\documentclass[longauth]{aa}

\usepackage{eht}
\usepackage[varg]{txfonts}
\usepackage{tabularx}
\usepackage{xcolor}
\usepackage[normalem]{ulem}
\usepackage{comment}
\usepackage{animate}
\usepackage{booktabs}
\usepackage{placeins}
\usepackage{needspace}

\usepackage{makecell}
\usepackage{siunitx}         
\usepackage{pifont}
\usepackage{colortbl}    
\usepackage{amssymb}     
\usepackage{multirow}    
\usepackage{float}       
\usepackage{threeparttable} 
\usepackage{placeins}    
\usepackage{enumitem} 

\definecolor{algkeyword}{RGB}{180,0,80}

\usepackage[symbol]{footmisc}

\usepackage{hyperref}
\hypersetup{
    colorlinks,
    linkcolor={blue!50!black},
    citecolor={blue!50!black},
    urlcolor={blue!50!black}
    }

\usepackage[encapsulated]{CJK}
\usepackage{ucs}
\usepackage[utf8]{inputenc} 
\newcommand{\cntext}[1]{\begin{CJK}{UTF8}{gbsn}#1\end{CJK}}

\usepackage{lmodern}
\newcommand{\orcid}[1]{}

\usepackage{subcaption}

\makeatletter

\begin{document}

\title{Dynamics and geometry of the inner sub-parsec-scale jet in 3C 279 
observed with the Event Horizon Telescope}
\titlerunning{Inner subpc-scale jet in 3C~279}

\author{Hendrik Müller\orcid{0000-0002-9250-0197}\inst{\ref{inst7},\ref{inst64}}\and
Sebastiano D. von Fellenberg\orcid{0000-0002-9156-2249}\inst{\ref{inst51},\ref{inst7}}\and
Ai-Ling Zeng (\cntext{曾艾玲})\orcid{0009-0000-9427-4608}\inst{\ref{inst6}}\and
Paul Tiede\orcid{0000-0003-3826-5648}\inst{\ref{inst4},\ref{inst12}}\and
Thomas P. Krichbaum\orcid{0000-0002-4892-9586}\inst{\ref{inst7}}\and
Roman Gold\orcid{0000-0003-2492-1966}\inst{\ref{inst75},\ref{inst76},\ref{inst77}}\and
Tuomas Savolainen\orcid{0000-0001-6214-1085}\inst{\ref{inst157},\ref{inst98},\ref{inst7}}\and
Jae-Young Kim\orcid{0000-0001-8229-7183}\inst{\ref{inst104}}\and
Sijia Peng \orcid{0000-0001-8492-892X}\inst{\ref{inst43}}
Teresa Toscano\orcid{0000-0003-3658-7862}\inst{\ref{inst6},\ref{inst7}}\and
Michael Janssen\orcid{0000-0001-8685-6544}\inst{\ref{inst32},\ref{inst7}}\and
Boris Georgiev\orcid{0000-0002-3586-6424}\inst{\ref{inst18}}\and
Dhanya G. Nair\orcid{0000-0001-5357-7805}\inst{\ref{inst19},\ref{inst7}}\and
Iniyan Natarajan\orcid{0000-0001-8242-4373}\inst{\ref{inst136},\ref{inst4},\ref{inst12}}\and
Lindy Blackburn\orcid{0000-0002-9030-642X}\inst{\ref{inst4},\ref{inst12}}\and
\\
The Event Horizon Telescope Collaboration
\\
Kazunori Akiyama\orcid{0000-0002-9475-4254}\inst{\ref{inst1},\ref{inst2},\ref{inst3},\ref{inst4}}\and
Ezequiel Albentosa-Ruíz\orcid{0000-0002-7816-6401}\inst{\ref{inst5}}\and
Antxon Alberdi\orcid{0000-0002-9371-1033}\inst{\ref{inst6}}\and
Walter Alef\inst{\ref{inst7}}\and
Juan Carlos Algaba\orcid{0000-0001-6993-1696}\inst{\ref{inst8}}\and
Rohan Ganesh Amanaganti\orcid{0009-0002-3431-4625}\inst{\ref{inst9}}\and
Richard Anantua\orcid{0000-0003-3457-7660}\inst{\ref{inst10},\ref{inst11},\ref{inst12},\ref{inst4}}\and
Eleni Antonopoulou\orcid{0009-0004-7747-7760}\inst{\ref{inst13},\ref{inst14}}\and
Keiichi Asada\orcid{0000-0001-6988-8763}\inst{\ref{inst15}}\and
Rebecca Azulay\orcid{0000-0002-2200-5393}\inst{\ref{inst5},\ref{inst16},\ref{inst7}}\and
Anne-Kathrin Baczko\orcid{0000-0003-3090-3975}\inst{\ref{inst17},\ref{inst7}}\and
David Ball\inst{\ref{inst18}}\and
Bidisha Bandyopadhyay\orcid{0000-0002-2138-8564}\inst{\ref{inst19}}\and
John Barrett\orcid{0000-0002-9290-0764}\inst{\ref{inst2}}\and
Michi Bauböck\orcid{0000-0002-5518-2812}\inst{\ref{inst20}}\and
Bradford A. Benson\orcid{0000-0002-5108-6823}\inst{\ref{inst21},\ref{inst22}}\and
Dan Bintley\inst{\ref{inst23},\ref{inst24}}\and
Raymond Blundell\orcid{0000-0002-5929-5857}\inst{\ref{inst4}}\and
Katherine L. Bouman\orcid{0000-0003-0077-4367}\inst{\ref{inst25}}\and
Geoffrey C. Bower\orcid{0000-0003-4056-9982}\inst{\ref{inst23},\ref{inst24},\ref{inst26},\ref{inst27}}\and
Michael Bremer\orcid{0000-0001-7511-3745}\inst{\ref{inst28}}\and
Roger Brissenden\orcid{0000-0002-2556-0894}\inst{\ref{inst4}}\and
Silke Britzen\orcid{0000-0001-9240-6734}\inst{\ref{inst7}}\and
Avery E. Broderick\orcid{0000-0002-3351-760X}\inst{\ref{inst29},\ref{inst30},\ref{inst31}}\and
Dominique Broguiere\orcid{0000-0001-9151-6683}\inst{\ref{inst28}}\and
Thomas Bronzwaer\orcid{0000-0003-1151-3971}\inst{\ref{inst32}}\and
Sandra Bustamante\orcid{0000-0001-6169-1894}\inst{\ref{inst33}}\and
Douglas F. Carlos\orcid{0000-0002-1340-7702}\inst{\ref{inst34}}\and
John E. Carlstrom\orcid{0000-0002-2044-7665}\inst{\ref{inst35},\ref{inst22},\ref{inst36},\ref{inst37}}\and
Andrew Chael\orcid{0000-0003-2966-6220}\inst{\ref{inst38}}\and
Chi-kwan Chan\orcid{0000-0001-6337-6126}\inst{\ref{inst18},\ref{inst39},\ref{inst40}}\and
Chin-Shin Chang\orcid{0000-0001-9910-3234}\inst{\ref{inst41}}\and
Dominic O. Chang\orcid{0000-0001-9939-5257}\inst{\ref{inst4},\ref{inst12}}\and
Koushik Chatterjee\orcid{0000-0002-2825-3590}\inst{\ref{inst42},\ref{inst12},\ref{inst4}}\and
Erandi Chavez\orcid{0000-0003-4143-9717}\inst{\ref{inst4}}\and
Ming-Tang Chen\orcid{0000-0001-6573-3318}\inst{\ref{inst26}}\and
Yongjun Chen (\cntext{陈永军})\orcid{0000-0001-5650-6770}\inst{\ref{inst43},\ref{inst44}}\and
Xiaopeng Cheng\orcid{0000-0003-4407-9868}\inst{\ref{inst45}}\and
Paul Chichura\orcid{0000-0002-5397-9035}\inst{\ref{inst36},\ref{inst35}}\and
Ilje Cho\orcid{0000-0001-6083-7521}\inst{\ref{inst46},\ref{inst45},\ref{inst6}}\and
Nicholas S. Conroy\orcid{0000-0003-2886-2377}\inst{\ref{inst47},\ref{inst4}}\and
John E. Conway\orcid{0000-0003-2448-9181}\inst{\ref{inst17}}\and
Thomas M. Crawford\orcid{0000-0001-9000-5013}\inst{\ref{inst22},\ref{inst35}}\and
Geoffrey B. Crew\orcid{0000-0002-2079-3189}\inst{\ref{inst2}}\and
Alejandro Cruz-Osorio\orcid{0000-0002-3945-6342}\inst{\ref{inst48}}\and
Yuzhu Cui (\cntext{崔玉竹})\orcid{0000-0001-6311-4345}\inst{\ref{inst49}}\and
Brandon Curd\orcid{0000-0002-8650-0879}\inst{\ref{inst10},\ref{inst12},\ref{inst4}}\and
Rohan Dahale\orcid{0000-0001-6982-9034}\inst{\ref{inst6},\ref{inst50},\ref{inst51}}\and
Jordy Davelaar\orcid{0000-0002-2685-2434}\inst{\ref{inst52},\ref{inst53}}\and
Joost de Kleuver\orcid{0009-0001-9624-1751}\inst{\ref{inst32}}\and
Mariafelicia De Laurentis\orcid{0000-0002-9945-682X}\inst{\ref{inst54},\ref{inst55}}\and
Roger Deane\orcid{0000-0003-1027-5043}\inst{\ref{inst56},\ref{inst57},\ref{inst58}}\and
Jason Dexter\orcid{0000-0003-3903-0373}\inst{\ref{inst59}}\and
Vedant Dhruv\orcid{0000-0001-6765-877X}\inst{\ref{inst20}}\and
Indu K. Dihingia\orcid{0000-0002-4064-0446}\inst{\ref{inst60},\ref{inst61}}\and
Sheperd S. Doeleman\orcid{0000-0002-9031-0904}\inst{\ref{inst4},\ref{inst12}}\and
Sergio A. Dzib\orcid{0000-0001-6010-6200}\inst{\ref{inst7}}\and
Razieh Emami\orcid{0000-0002-2791-5011}\inst{\ref{inst4}}\and
Heino Falcke\orcid{0000-0002-2526-6724}\inst{\ref{inst32}}\and
Joseph Farah\orcid{0000-0003-4914-5625}\inst{\ref{inst62},\ref{inst63}}\and
Vincent L. Fish\orcid{0000-0002-7128-9345}\inst{\ref{inst2}}\and
Edward Fomalont\orcid{0000-0002-9036-2747}\inst{\ref{inst64}}\and
H. Alyson Ford\orcid{0000-0002-9797-0972}\inst{\ref{inst18}}\and
Marianna Foschi\orcid{0000-0001-8147-4993}\inst{\ref{inst6},\ref{inst25}}\and
Raquel Fraga-Encinas\orcid{0000-0002-5222-1361}\inst{\ref{inst32}}\and
William T. Freeman\inst{\ref{inst65},\ref{inst66}}\and
Per Friberg\orcid{0000-0002-8010-8454}\inst{\ref{inst23},\ref{inst24}}\and
Christian M. Fromm\orcid{0000-0002-1827-1656}\inst{\ref{inst67},\ref{inst68},\ref{inst7}}\and
Antonio Fuentes\orcid{0000-0002-8773-4933}\inst{\ref{inst6}}\and
Peter Galison\orcid{0000-0002-6429-3872}\inst{\ref{inst12},\ref{inst69},\ref{inst70}}\and
Charles F. Gammie\orcid{0000-0001-7451-8935}\inst{\ref{inst20},\ref{inst47},\ref{inst71}}\and
Roberto García\orcid{0000-0002-6584-7443}\inst{\ref{inst28}}\and
Olivier Gentaz\orcid{0000-0002-0115-4605}\inst{\ref{inst28}}\and
Ciriaco Goddi\orcid{0000-0002-2542-7743}\inst{\ref{inst34},\ref{inst72},\ref{inst73},\ref{inst74}}\and
Arturo I. Gómez-Ruiz\orcid{0000-0001-9395-1670}\inst{\ref{inst78},\ref{inst79}}\and
Brissa Gomez Miller\orcid{0009-0006-8345-6805}\inst{\ref{inst80}}\and
José L. Gómez\orcid{0000-0003-4190-7613}\inst{\ref{inst6}}\and
Minfeng Gu (\cntext{顾敏峰})\orcid{0000-0002-4455-6946}\inst{\ref{inst43},\ref{inst81}}\and
Mark Gurwell\orcid{0000-0003-0685-3621}\inst{\ref{inst4}}\and
Kazuhiro Hada\orcid{0000-0001-6906-772X}\inst{\ref{inst82},\ref{inst3}}\and
Daryl Haggard\orcid{0000-0001-6803-2138}\inst{\ref{inst83},\ref{inst84}}\and
Ronald Hesper\orcid{0000-0003-1918-6098}\inst{\ref{inst85}}\and
Dirk Heumann\orcid{0000-0002-7671-0047}\inst{\ref{inst18}}\and
Luis C. Ho (\cntext{何子山})\orcid{0000-0001-6947-5846}\inst{\ref{inst86},\ref{inst87}}\and
Paul Ho\orcid{0000-0002-3412-4306}\inst{\ref{inst15},\ref{inst24},\ref{inst23}}\and
Mareki Honma\orcid{0000-0003-4058-9000}\inst{\ref{inst3},\ref{inst88},\ref{inst89}}\and
Chih-Wei L. Huang\orcid{0000-0001-5641-3953}\inst{\ref{inst15}}\and
Lei Huang (\cntext{黄磊})\orcid{0000-0002-1923-227X}\inst{\ref{inst43},\ref{inst81}}\and
David H. Hughes\inst{\ref{inst78}}\and
Shiro Ikeda\orcid{0000-0002-2462-1448}\inst{\ref{inst90},\ref{inst91},\ref{inst92},\ref{inst93}}\and
C. M. Violette Impellizzeri\orcid{0000-0002-3443-2472}\inst{\ref{inst94},\ref{inst64}}\and
Makoto Inoue\orcid{0000-0001-5037-3989}\inst{\ref{inst15}}\and
Sara Issaoun\orcid{0000-0002-5297-921X}\inst{\ref{inst4},\ref{inst53}}\and
Yuhei Iwata\orcid{0000-0002-9255-4742}\inst{\ref{inst3},\ref{inst88}}\and
David J. James\orcid{0000-0001-5160-4486}\inst{\ref{inst95},\ref{inst96}}\and
Buell T. Jannuzi\orcid{0000-0002-1578-6582}\inst{\ref{inst18}}\and
Britton Jeter\orcid{0000-0003-2847-1712}\inst{\ref{inst97},\ref{inst98}}\and
Wu Jiang (\cntext{江悟})\orcid{0000-0001-7369-3539}\inst{\ref{inst43}}\and
Alejandra Jiménez-Rosales\orcid{0000-0002-2662-3754}\inst{\ref{inst32}}\and
Michael D. Johnson\orcid{0000-0002-4120-3029}\inst{\ref{inst4},\ref{inst12}}\and
Svetlana Jorstad\orcid{0000-0001-6158-1708}\inst{\ref{inst99}}\and
Adam C. Jones\inst{\ref{inst22}}\and
Abhishek V. Joshi\orcid{0000-0002-2514-5965}\inst{\ref{inst20}}\and
Taehyun Jung\orcid{0000-0001-7003-8643}\inst{\ref{inst45},\ref{inst100}}\and
Tomohisa Kawashima\orcid{0000-0001-8527-0496}\inst{\ref{inst101}}\and
Garrett K. Keating\orcid{0000-0002-3490-146X}\inst{\ref{inst4}}\and
Mark Kettenis\orcid{0000-0002-6156-5617}\inst{\ref{inst102}}\and
Dong-Jin Kim\orcid{0000-0002-7038-2118}\inst{\ref{inst103}}\and
Jongsoo Kim\orcid{0000-0002-1229-0426}\inst{\ref{inst45}}\and
Junhan Kim\orcid{0000-0002-4274-9373}\inst{\ref{inst105}}\and
Motoki Kino\orcid{0000-0002-2709-7338}\inst{\ref{inst90},\ref{inst106}}\and
Jakob Knollmüller\orcid{0000-0002-9906-0040}\inst{\ref{inst107},\ref{inst32}}\and
Jun Yi Koay\orcid{0000-0002-7029-6658}\inst{\ref{inst108},\ref{inst15}}\and
Prashant Kocherlakota\orcid{0000-0001-7386-7439}\inst{\ref{inst12},\ref{inst4}}\and
Yutaro Kofuji\inst{\ref{inst6},\ref{inst3}}\and
Patrick M. Koch\orcid{0000-0003-2777-5861}\inst{\ref{inst15}}\and
Shoko Koyama\orcid{0000-0002-3723-3372}\inst{\ref{inst108},\ref{inst15}}\and
Carsten Kramer\orcid{0000-0002-4908-4925}\inst{\ref{inst28}}\and
Joana A. Kramer\orcid{0009-0003-3011-0454}\inst{\ref{inst109},\ref{inst7}}\and
Michael Kramer\orcid{0000-0002-4175-2271}\inst{\ref{inst7}}\and
Cheng-Yu Kuo\orcid{0000-0001-6211-5581}\inst{\ref{inst110},\ref{inst15}}\and
Noemi La Bella\orcid{0000-0002-8116-9427}\inst{\ref{inst32}}\and
Deokhyeong Lee\orcid{0009-0003-2122-9437}\inst{\ref{inst46}}\and
Sang-Sung Lee\orcid{0000-0002-6269-594X}\inst{\ref{inst45}}\and
Aviad Levis\orcid{0000-0001-7307-632X}\inst{\ref{inst50},\ref{inst111},\ref{inst112}}\and
Shaoliang Li\orcid{0009-0005-0338-9490}\inst{\ref{inst23},\ref{inst24}}\and
Zhiyuan Li (\cntext{李志远})\orcid{0000-0003-0355-6437}\inst{\ref{inst113},\ref{inst114}}\and
Rocco Lico\orcid{0000-0001-7361-2460}\inst{\ref{inst115},\ref{inst6}}\and
Greg Lindahl\orcid{0000-0002-6100-4772}\inst{\ref{inst116}}\and
Michael Lindqvist\orcid{0000-0002-3669-0715}\inst{\ref{inst17}}\and
Mikhail Lisakov\orcid{0000-0001-6088-3819}\inst{\ref{inst117}}\and
Jun Liu (\cntext{刘俊})\orcid{0000-0002-7615-7499}\inst{\ref{inst7}}\and
Kuo Liu\orcid{0000-0002-2953-7376}\inst{\ref{inst118}}\and
Elisabetta Liuzzo\orcid{0000-0003-0995-5201}\inst{\ref{inst119}}\and
Wen-Ping Lo\orcid{0000-0003-1869-2503}\inst{\ref{inst15},\ref{inst120}}\and
Andrei P. Lobanov\orcid{0000-0003-1622-1484}\inst{\ref{inst7}}\and
Laurent Loinard\orcid{0000-0002-5635-3345}\inst{\ref{inst80},\ref{inst12}}\and
Colin J. Lonsdale\orcid{0000-0003-4062-4654}\inst{\ref{inst2}}\and
Amy E. Lowitz\orcid{0000-0002-4747-4276}\inst{\ref{inst18}}\and
Ru-Sen Lu (\cntext{路如森})\orcid{0000-0002-7692-7967}\inst{\ref{inst43},\ref{inst44},\ref{inst7}}\and
Nicholas R. MacDonald\orcid{0000-0002-6684-8691}\inst{\ref{inst121},\ref{inst7}}\and
Jirong Mao (\cntext{毛基荣})\orcid{0000-0002-7077-7195}\inst{\ref{inst122},\ref{inst123},\ref{inst124}}\and
Nicola Marchili\orcid{0000-0002-5523-7588}\inst{\ref{inst119},\ref{inst7}}\and
Sera Markoff\orcid{0000-0001-9564-0876}\inst{\ref{inst109},\ref{inst125},\ref{inst126}}\and
Daniel P. Marrone\orcid{0000-0002-2367-1080}\inst{\ref{inst18}}\and
Alan P. Marscher\orcid{0000-0001-7396-3332}\inst{\ref{inst99}}\and
Iván Martí-Vidal\orcid{0000-0003-3708-9611}\inst{\ref{inst5},\ref{inst16}}\and
Satoki Matsushita\orcid{0000-0002-2127-7880}\inst{\ref{inst15}}\and
Lynn D. Matthews\orcid{0000-0002-3728-8082}\inst{\ref{inst2}}\and
Lia Medeiros\orcid{0000-0003-2342-6728}\inst{\ref{inst9}}\and
Karl M. Menten\orcid{0000-0001-6459-0669}\inst{\ref{inst7},\ref{inst127}}\and
Hugo Messias\orcid{0000-0002-2985-7994}\inst{\ref{inst128},\ref{inst129}}\and
Izumi Mizuno\orcid{0000-0002-7210-6264}\inst{\ref{inst23},\ref{inst24}}\and
Yosuke Mizuno\orcid{0000-0002-8131-6730}\inst{\ref{inst61},\ref{inst130},\ref{inst68}}\and
Joshua Montgomery\orcid{0000-0003-0345-8386}\inst{\ref{inst84},\ref{inst22}}\and
Kotaro Moriyama\orcid{0000-0003-1364-3761}\inst{\ref{inst6},\ref{inst3}}\and
Monika Moscibrodzka\orcid{0000-0002-4661-6332}\inst{\ref{inst32}}\and
Wanga Mulaudzi\orcid{0000-0003-4514-625X}\inst{\ref{inst109}}\and
Cornelia Müller\orcid{0000-0002-2739-2994}\inst{\ref{inst7},\ref{inst32}}\and
Alejandro Mus\orcid{0000-0003-0329-6874}\inst{\ref{inst72},\ref{inst115},\ref{inst131},\ref{inst132},\ref{inst133}}\and
Gibwa Musoke\orcid{0000-0003-1984-189X}\inst{\ref{inst109},\ref{inst32}}\and
Ioannis Myserlis\orcid{0000-0003-3025-9497}\inst{\ref{inst134}}\and
Hiroshi Nagai\orcid{0000-0003-0292-3645}\inst{\ref{inst90},\ref{inst88}}\and
Neil M. Nagar\orcid{0000-0001-6920-662X}\inst{\ref{inst19}}\and
Masanori Nakamura\orcid{0000-0001-6081-2420}\inst{\ref{inst135},\ref{inst15}}\and
Gopal Narayanan\orcid{0000-0002-4723-6569}\inst{\ref{inst33}}\and
Antonios Nathanail\orcid{0000-0002-1655-9912}\inst{\ref{inst13}}\and
Santiago Navarro Fuentes\inst{\ref{inst134}}\and
Joey Neilsen\orcid{0000-0002-8247-786X}\inst{\ref{inst137}}\and
Chunchong Ni\orcid{0000-0003-1361-5699}\inst{\ref{inst30},\ref{inst31},\ref{inst29}}\and
Andy Nilipour\orcid{0000-0002-5956-5167}\inst{\ref{inst138},\ref{inst139}}\and
Michael A. Nowak\orcid{0000-0001-6923-1315}\inst{\ref{inst140}}\and
Hiroki Okino\orcid{0000-0003-3779-2016}\inst{\ref{inst3},\ref{inst89}}\and
Héctor Raúl Olivares Sánchez\orcid{0000-0001-6833-7580}\inst{\ref{inst141}}\and
Feryal Özel\orcid{0000-0003-4413-1523}\inst{\ref{inst142}}\and
Daniel C. M. Palumbo\orcid{0000-0002-7179-3816}\inst{\ref{inst12},\ref{inst4}}\and
Georgios Filippos Paraschos\orcid{0000-0001-6757-3098}\inst{\ref{inst97},\ref{inst98},\ref{inst7}}\and
Jongho Park\orcid{0000-0001-6558-9053}\inst{\ref{inst143},\ref{inst144},\ref{inst15}}\and
Harriet Parsons\orcid{0000-0002-6327-3423}\inst{\ref{inst23},\ref{inst24}}\and
Nimesh Patel\orcid{0000-0002-6021-9421}\inst{\ref{inst4}}\and
Ue-Li Pen\orcid{0000-0003-2155-9578}\inst{\ref{inst15},\ref{inst29},\ref{inst51},\ref{inst112},\ref{inst145}}\and
Dominic W. Pesce\orcid{0000-0002-5278-9221}\inst{\ref{inst4},\ref{inst12}}\and
Vincent Piétu\orcid{0009-0006-3497-397X}\inst{\ref{inst28}}\and
Alexander Plavin\orcid{0000-0003-2914-8554}\inst{\ref{inst12},\ref{inst4},\ref{inst7}}\and
Aleksandar PopStefanija\inst{\ref{inst33}}\and
Oliver Porth\orcid{0000-0002-4584-2557}\inst{\ref{inst109},\ref{inst68}}\and
Cora Prather\orcid{0000-0002-0393-7734}\inst{\ref{inst12}}\and
Giacomo Principe\orcid{0000-0003-0406-7387}\inst{\ref{inst146},\ref{inst147},\ref{inst115}}\and
Dimitrios Psaltis\orcid{0000-0003-1035-3240}\inst{\ref{inst142}}\and
Hung-Yi Pu\orcid{0000-0001-9270-8812}\inst{\ref{inst148},\ref{inst149},\ref{inst15}}\and
Alexandra Rahlin\orcid{0000-0003-3953-1776}\inst{\ref{inst22}}\and
Venkatessh Ramakrishnan\orcid{0000-0002-9248-086X}\inst{\ref{inst150},\ref{inst97},\ref{inst98}}\and
Ramprasad Rao\orcid{0000-0002-1407-7944}\inst{\ref{inst4}}\and
Mark G. Rawlings\orcid{0000-0002-6529-202X}\inst{\ref{inst64},\ref{inst23},\ref{inst24}}\and
Luciano Rezzolla\orcid{0000-0002-1330-7103}\inst{\ref{inst68},\ref{inst151},\ref{inst152}}\and
Angelo Ricarte\orcid{0000-0001-5287-0452}\inst{\ref{inst12},\ref{inst4}}\and
Luca Ricci\orcid{0000-0002-4175-3194}\inst{\ref{inst153}}\and
Bart Ripperda\orcid{0000-0002-7301-3908}\inst{\ref{inst51},\ref{inst154},\ref{inst112},\ref{inst29}}\and
Jan Röder\orcid{0000-0002-2426-927X}\inst{\ref{inst6}}\and
Freek Roelofs\orcid{0000-0001-5461-3687}\inst{\ref{inst32}}\and
Cristina Romero-Cañizales\orcid{0000-0001-6301-9073}\inst{\ref{inst15}}\and
Eduardo Ros\orcid{0000-0001-9503-4892}\inst{\ref{inst7}}\and
Arash Roshanineshat\orcid{0000-0002-8280-9238}\inst{\ref{inst18}}\and
Helge Rottmann\inst{\ref{inst7}}\and
Alan L. Roy\orcid{0000-0002-1931-0135}\inst{\ref{inst7}}\and
Ignacio Ruiz\orcid{0000-0002-0965-5463}\inst{\ref{inst134}}\and
Chet Ruszczyk\orcid{0000-0001-7278-9707}\inst{\ref{inst2}}\and
Kazi L. J. Rygl\orcid{0000-0003-4146-9043}\inst{\ref{inst119}}\and
León D. S. Salas\orcid{0000-0003-1979-6363}\inst{\ref{inst109}}\and
Salvador Sánchez\orcid{0000-0002-8042-5951}\inst{\ref{inst134}}\and
David Sánchez-Argüelles\orcid{0000-0002-7344-9920}\inst{\ref{inst78},\ref{inst79}}\and
Miguel Sánchez-Portal\orcid{0000-0003-0981-9664}\inst{\ref{inst134}}\and
Ali SaraerToosi\orcid{0009-0003-4620-8448}\inst{\ref{inst50}}\and
Mahito Sasada\orcid{0000-0001-5946-9960}\inst{\ref{inst155},\ref{inst3},\ref{inst156}}\and
Kaushik Satapathy\orcid{0000-0003-0433-3585}\inst{\ref{inst18}}\and
Saurabh\orcid{0000-0001-7156-4848}\inst{\ref{inst7}}\and
Karl-Friedrich Schuster\orcid{0000-0003-2890-9454}\inst{\ref{inst158}}\and
Zhiqiang Shen (\cntext{沈志强})\orcid{0000-0003-3540-8746}\inst{\ref{inst43},\ref{inst44}}\and
Sasikumar Silpa\orcid{0000-0003-0667-7074}\inst{\ref{inst19}}\and
Randall Smith\orcid{0000-0003-4284-4167}\inst{\ref{inst4}}\and
Bong Won Sohn\orcid{0000-0002-4148-8378}\inst{\ref{inst45},\ref{inst100}}\and
Jason SooHoo\orcid{0000-0003-1938-0720}\inst{\ref{inst2}}\and
Kamal Souccar\orcid{0000-0001-7915-5272}\inst{\ref{inst33}}\and
Joshua S. Stanway\orcid{0009-0003-7659-4642}\inst{\ref{inst159}}\and
He Sun (\cntext{孙赫})\orcid{0000-0003-1526-6787}\inst{\ref{inst160},\ref{inst161}}\and
Alexandra J. Tetarenko\orcid{0000-0003-3906-4354}\inst{\ref{inst162}}\and
Remo P. J. Tilanus\orcid{0000-0002-6514-553X}\inst{\ref{inst18}}\and
Michael Titus\orcid{0000-0001-9001-3275}\inst{\ref{inst2}}\and
Kenji Toma\orcid{0000-0002-7114-6010}\inst{\ref{inst163},\ref{inst164}}\and
Pablo Torne\orcid{0000-0001-8700-6058}\inst{\ref{inst134},\ref{inst7}}\and
Efthalia Traianou\orcid{0000-0002-1209-6500}\inst{\ref{inst6},\ref{inst7}}\and
Sascha Trippe\orcid{0000-0003-0465-1559}\inst{\ref{inst165},\ref{inst166}}\and
Matthew Turk\orcid{0000-0002-5294-0198}\inst{\ref{inst47}}\and
Akhil Uniyal\orcid{0000-0001-8213-646X}\inst{\ref{inst61}}\and
Ilse van Bemmel\orcid{0000-0001-5473-2950}\inst{\ref{inst167}}\and
Bram van den Berg\orcid{0009-0000-9340-4204}\inst{\ref{inst32}}\and
Huib Jan van Langevelde\orcid{0000-0002-0230-5946}\inst{\ref{inst102},\ref{inst94}}\and
Daniel R. van Rossum\orcid{0000-0001-7772-6131}\inst{\ref{inst32}}\and
Jesse Vos\orcid{0000-0003-3349-7394}\inst{\ref{inst168}}\and
Jan Wagner\orcid{0000-0003-1105-6109}\inst{\ref{inst7}}\and
Zhiren Wang\orcid{0009-0004-9417-2213}\inst{\ref{inst29},\ref{inst30},\ref{inst31}}\and
Derek Ward-Thompson\orcid{0000-0003-1140-2761}\inst{\ref{inst159}}\and
John Wardle\orcid{0000-0002-8960-2942}\inst{\ref{inst169}}\and
Jasmin E. Washington\orcid{0000-0002-7046-0470}\inst{\ref{inst18}}\and
Jonathan Weintroub\orcid{0000-0002-4603-5204}\inst{\ref{inst4},\ref{inst12}}\and
Maciek Wielgus\orcid{0000-0002-8635-4242}\inst{\ref{inst6}}\and
Kaj Wiik\orcid{0000-0002-0862-3398}\inst{\ref{inst170},\ref{inst97},\ref{inst98}}\and
Michael F. Wondrak\orcid{0000-0002-6894-1072}\inst{\ref{inst109},\ref{inst125},\ref{inst32},\ref{inst171}}\and
George N. Wong\orcid{0000-0001-6952-2147}\inst{\ref{inst172},\ref{inst38}}\and
Jompoj Wongphexhauxsorn\orcid{0000-0002-7730-4956}\inst{\ref{inst153},\ref{inst7}}\and
Qingwen Wu (\cntext{吴庆文})\orcid{0000-0003-4773-4987}\inst{\ref{inst173}}\and
Paul Yamaguchi\orcid{0000-0002-6017-8199}\inst{\ref{inst4}}\and
Aristomenis Yfantis\orcid{0000-0002-3244-7072}\inst{\ref{inst6}}\and
Doosoo Yoon\orcid{0000-0001-8694-8166}\inst{\ref{inst109}}\and
André Young\orcid{0000-0003-0000-2682}\inst{\ref{inst32}}\and
Ziri Younsi\orcid{0000-0001-9283-1191}\inst{\ref{inst174},\ref{inst68}}\and
Wei Yu (\cntext{于威})\orcid{0000-0002-5168-6052}\inst{\ref{inst4}}\and
Feng Yuan (\cntext{袁峰})\orcid{0000-0003-3564-6437}\inst{\ref{inst175}}\and
Ye-Fei Yuan (\cntext{袁业飞})\orcid{0000-0002-7330-4756}\inst{\ref{inst176}}\and
J. Anton Zensus\orcid{0000-0001-7470-3321}\inst{\ref{inst7}}\and
Shuo Zhang\orcid{0000-0002-2967-790X}\inst{\ref{inst177}}\and
Brandon Zhao\orcid{0009-0005-3991-9879}\inst{\ref{inst25}}\and
Guang-Yao Zhao\orcid{0000-0002-4417-1659}\inst{\ref{inst7},\ref{inst6}}
}
\institute{
Max-Planck-Institut für Radioastronomie, Auf dem Hügel 69, D-53121 Bonn, Germany\label{inst7}\and
National Radio Astronomy Observatory, 520 Edgemont Road, Charlottesville, VA 22903, USA\label{inst64}\and
Institute of Sensors, Signals and Systems, Heriot-Watt University, Edinburgh EH14 4AS, United Kingdom\label{inst1}\and
Massachusetts Institute of Technology Haystack Observatory, 99 Millstone Road, Westford, MA 01886, USA\label{inst2}\and
Mizusawa VLBI Observatory, National Astronomical Observatory of Japan, 2-12 Hoshigaoka, Mizusawa, Oshu, Iwate 023-0861, Japan\label{inst3}\and
Center for Astrophysics $|$ Harvard \& Smithsonian, 60 Garden Street, Cambridge, MA 02138, USA\label{inst4}\and
Departament d'Astronomia i Astrofísica, Universitat de València, C. Dr. Moliner 50, E-46100 Burjassot, València, Spain\label{inst5}\and
Instituto de Astrofísica de Andalucía-CSIC, Glorieta de la Astronomía s/n, E-18008 Granada, Spain\label{inst6}\and
Centre for Astronomy and Astrophysics Research, Department of Physics, Faculty of Science, Universiti Malaya, 50603 Kuala Lumpur, Malaysia\label{inst8}\and
Center for Gravitation, Cosmology and Astrophysics, Department of Physics, University of Wisconsin–Milwaukee, P.O. Box 413, Milwaukee, WI 53201, USA\label{inst9}\and
Department of Physics \& Astronomy, The University of Texas at San Antonio, One UTSA Circle, San Antonio, TX 78249, USA\label{inst10}\and
Physics \& Astronomy Department, Rice University, Houston, TX 77005-1827, USA\label{inst11}\and
Black Hole Initiative at Harvard University, 20 Garden Street, Cambridge, MA 02138, USA\label{inst12}\and
Research Center for Astronomy, Academy of Athens, Soranou Efessiou 4, 115 27 Athens, Greece\label{inst13}\and
Department of Physics, National and Kapodistrian University of Athens, Panepistimiopolis, GR 15783 Zografos, Greece\label{inst14}\and
Institute of Astronomy and Astrophysics, Academia Sinica, 11F of Astronomy-Mathematics Building, AS/NTU No. 1, Sec. 4, Roosevelt Rd., Taipei 106216, Taiwan, R.O.C.\label{inst15}\and
Observatori Astronòmic, Universitat de València, C. Catedrático José Beltrán 2, E-46980 Paterna, València, Spain\label{inst16}\and
Department of Physics and Astronomy, Chalmers University of Technology, Onsala Space Observatory, SE-439 92 Onsala, Sweden\label{inst17}\and
Steward Observatory and Department of Astronomy, University of Arizona, 933 N. Cherry Ave., Tucson, AZ 85721, USA\label{inst18}\and
Astronomy Department, Universidad de Concepción, Casilla 160-C, Concepción, Chile\label{inst19}\and
Department of Physics, University of Illinois, 1110 West Green Street, Urbana, IL 61801, USA\label{inst20}\and
Fermi National Accelerator Laboratory, MS209, P.O. Box 500, Batavia, IL 60510, USA\label{inst21}\and
Department of Astronomy and Astrophysics, University of Chicago, 5640 South Ellis Avenue, Chicago, IL 60637, USA\label{inst22}\and
East Asian Observatory, 660 N. A'ohoku Place, Hilo, HI 96720, USA\label{inst23}\and
James Clerk Maxwell Telescope (JCMT), 660 N. A'ohoku Place, Hilo, HI 96720, USA\label{inst24}\and
California Institute of Technology, 1200 East California Boulevard, Pasadena, CA 91125, USA\label{inst25}\and
Institute of Astronomy and Astrophysics, Academia Sinica, 645 N. A'ohoku Place, Hilo, HI 96720, USA\label{inst26}\and
Department of Physics and Astronomy, University of Hawaii at Manoa, 2505 Correa Road, Honolulu, HI 96822, USA\label{inst27}\and
Institut de Radioastronomie Millimétrique (IRAM), 300 rue de la Piscine, F-38400 Saint-Martin-d'Hères, France\label{inst28}\and
Perimeter Institute for Theoretical Physics, 31 Caroline Street North, Waterloo, ON N2L 2Y5, Canada\label{inst29}\and
Department of Physics and Astronomy, University of Waterloo, 200 University Avenue West, Waterloo, ON N2L 3G1, Canada\label{inst30}\and
Waterloo Centre for Astrophysics, University of Waterloo, Waterloo, ON N2L 3G1, Canada\label{inst31}\and
Department of Astrophysics, Institute for Mathematics, Astrophysics and Particle Physics (IMAPP), Radboud University, P.O. Box 9010, 6500 GL Nijmegen, The Netherlands\label{inst32}\and
Department of Astronomy, University of Massachusetts, Amherst, MA 01003, USA\label{inst33}\and
Instituto de Astronomia, Geofísica e Ciências Atmosféricas, Universidade de São Paulo, R. do Matão, 1226, São Paulo, SP 05508-090, Brazil\label{inst34}\and
Kavli Institute for Cosmological Physics, University of Chicago, 5640 South Ellis Avenue, Chicago, IL 60637, USA\label{inst35}\and
Department of Physics, University of Chicago, 5720 South Ellis Avenue, Chicago, IL 60637, USA\label{inst36}\and
Enrico Fermi Institute, University of Chicago, 5640 South Ellis Avenue, Chicago, IL 60637, USA\label{inst37}\and
Princeton Gravity Initiative, Jadwin Hall, Princeton University, Princeton, NJ 08544, USA\label{inst38}\and
Data Science Institute, University of Arizona, 1230 N. Cherry Ave., Tucson, AZ 85721, USA\label{inst39}\and
Program in Applied Mathematics, University of Arizona, 617 N. Santa Rita, Tucson, AZ 85721, USA\label{inst40}\and
Department of Astronomy, University of Geneva, Chemin Pegasi 51, 1290 Versoix, Switzerland\label{inst41}\and
Department of Physics, University of Maryland, 7901 Regents Drive, College Park, MD 20742, USA\label{inst42}\and
Shanghai Astronomical Observatory, Chinese Academy of Sciences, 80 Nandan Road, Shanghai 200030, People's Republic of China\label{inst43}\and
Key Laboratory of Radio Astronomy and Technology, Chinese Academy of Sciences, A20 Datun Road, Chaoyang District, Beijing, 100101, People’s Republic of China\label{inst44}\and
Korea Astronomy and Space Science Institute, Daedeok-daero 776, Yuseong-gu, Daejeon 34055, Republic of Korea\label{inst45}\and
Department of Astronomy, Kyungpook National University, 80 Daehak-ro, Buk-gu, Daegu 41566, Republic of Korea\label{inst46}\and
Department of Astronomy, University of Illinois at Urbana-Champaign, 1002 West Green Street, Urbana, IL 61801, USA\label{inst47}\and
Instituto de Astronomía, Universidad Nacional Autónoma de México (UNAM), Apdo Postal 70-264, Ciudad de México, México\label{inst48}\and
Institute of Astrophysics, Central China Normal University, Wuhan 430079, People's Republic of China\label{inst49}\and
Department of Computer Science, University of Toronto, 40 St. George St., Toronto, ON, M5S 2E4, Canada\label{inst50}\and
Canadian Institute for Theoretical Astrophysics, University of Toronto, 60 St. George Street, Toronto, ON M5S 3H8, Canada\label{inst51}\and
Department of Astrophysical Sciences, Peyton Hall, Princeton University, Princeton, NJ 08544, USA\label{inst52}\and
NASA Hubble Fellowship Program, Einstein Fellow\label{inst53}\and
Dipartimento di Fisica ``E. Pancini'', Università di Napoli ``Federico II'', Compl. Univ. di Monte S. Angelo, Edificio G, Via Cinthia, I-80126, Napoli, Italy\label{inst54}\and
INFN Sez. di Napoli, Compl. Univ. di Monte S. Angelo, Edificio G, Via Cinthia, I-80126, Napoli, Italy\label{inst55}\and
Wits Centre for Astrophysics, University of the Witwatersrand, 1 Jan Smuts Avenue, Braamfontein, Johannesburg 2050, South Africa\label{inst56}\and
Department of Physics, University of Pretoria, Hatfield, Pretoria 0028, South Africa\label{inst57}\and
Centre for Radio Astronomy Techniques and Technologies, Department of Physics and Electronics, Rhodes University, Makhanda 6140, South Africa\label{inst58}\and
JILA and Department of Astrophysical and Planetary Sciences, University of Colorado, Boulder, CO 80309, USA\label{inst59}\and
Institute of Fundamental Physics and Quantum Technology, \& School of Physical Science and Technology, Ningbo University, Ningbo, Zhejiang 315211, People’s Republic of China\label{inst60}\and
Tsung-Dao Lee Institute, Shanghai Jiao Tong University, Shengrong Road 520, Shanghai, 201210, People’s Republic of China\label{inst61}\and
Las Cumbres Observatory, 6740 Cortona Drive, Suite 102, Goleta, CA 93117-5575, USA\label{inst62}\and
Department of Physics, University of California, Santa Barbara, CA 93106-9530, USA\label{inst63}\and
Department of Electrical Engineering and Computer Science, Massachusetts Institute of Technology, 32-D476, 77 Massachusetts Ave., Cambridge, MA 02142, USA\label{inst65}\and
Google Research, 355 Main St., Cambridge, MA 02142, USA\label{inst66}\and
Institut für Theoretische Physik und Astrophysik, Universität Würzburg, Emil-Fischer-Str. 31, D-97074 Würzburg, Germany\label{inst67}\and
Institut für Theoretische Physik, Goethe-Universität Frankfurt, Max-von-Laue-Straße 1, D-60438 Frankfurt am Main, Germany\label{inst68}\and
Department of History of Science, Harvard University, Cambridge, MA 02138, USA\label{inst69}\and
Department of Physics, Harvard University, Cambridge, MA 02138, USA\label{inst70}\and
NCSA, University of Illinois, 1205 W. Clark St., Urbana, IL 61801, USA\label{inst71}\and
Dipartimento di Fisica, Università degli Studi di Cagliari, SP Monserrato-Sestu km 0.7, I-09042 Monserrato (CA), Italy\label{inst72}\and
INAF - Osservatorio Astronomico di Cagliari, via della Scienza 5, I-09047 Selargius (CA), Italy\label{inst73}\and
INFN, sezione di Cagliari, I-09042 Monserrato (CA), Italy\label{inst74}\and
Institute for Mathematics and Interdisciplinary Center for Scientific Computing, Heidelberg University, Im Neuenheimer Feld 205, Heidelberg 69120, Germany\label{inst75}\and
Institut f\"ur Theoretische Physik, Universit\"at Heidelberg, Philosophenweg 16, 69120 Heidelberg, Germany\label{inst76}\and
CP3-Origins, University of Southern Denmark, Campusvej 55, DK-5230 Odense, Denmark\label{inst77}\and
Instituto Nacional de Astrofísica, Óptica y Electrónica. Apartado Postal 51 y 216, 72000. Puebla Pue., México\label{inst78}\and
Consejo Nacional de Humanidades, Ciencia y Tecnología, Av. Insurgentes Sur 1582, 03940, Ciudad de México, México\label{inst79}\and
Instituto de Radioastronomía y Astrofísica, Universidad Nacional Autónoma de México, Morelia 58089, México\label{inst80}\and
Key Laboratory for Research in Galaxies and Cosmology, Chinese Academy of Sciences, Shanghai 200030, People's Republic of China\label{inst81}\and
Graduate School of Science, Nagoya City University, Yamanohata 1, Mizuho-cho, Mizuho-ku, Nagoya, 467-8501, Aichi, Japan\label{inst82}\and
Department of Physics, McGill University, 3600 rue University, Montréal, QC H3A 2T8, Canada\label{inst83}\and
Trottier Space Institute at McGill, 3550 rue University, Montréal,  QC H3A 2A7, Canada\label{inst84}\and
NOVA Sub-mm Instrumentation Group, Kapteyn Astronomical Institute, University of Groningen, Landleven 12, 9747 AD Groningen, The Netherlands\label{inst85}\and
Department of Astronomy, School of Physics, Peking University, Beijing 100871, People's Republic of China\label{inst86}\and
Kavli Institute for Astronomy and Astrophysics, Peking University, Beijing 100871, People's Republic of China\label{inst87}\and
Department of Astronomical Science, The Graduate University for Advanced Studies (SOKENDAI), 2-21-1 Osawa, Mitaka, Tokyo 181-8588, Japan\label{inst88}\and
Department of Astronomy, Graduate School of Science, The University of Tokyo, 7-3-1 Hongo, Bunkyo-ku, Tokyo 113-0033, Japan\label{inst89}\and
National Astronomical Observatory of Japan, 2-21-1 Osawa, Mitaka, Tokyo 181-8588, Japan\label{inst90}\and
The Institute of Statistical Mathematics, 10-3 Midori-cho, Tachikawa, Tokyo, 190-8562, Japan\label{inst91}\and
Department of Statistical Science, The Graduate University for Advanced Studies (SOKENDAI), 10-3 Midori-cho, Tachikawa, Tokyo 190-8562, Japan\label{inst92}\and
Kavli Institute for the Physics and Mathematics of the Universe, The University of Tokyo, 5-1-5 Kashiwanoha, Kashiwa, 277-8583, Japan\label{inst93}\and
Leiden Observatory, Leiden University, Postbus 2300, 9513 RA Leiden, The Netherlands\label{inst94}\and
ASTRAVEO LLC, PO Box 1668, Gloucester, MA 01931, USA\label{inst95}\and
Applied Materials Inc., 35 Dory Road, Gloucester, MA 01930, USA\label{inst96}\and
Finnish Centre for Astronomy with ESO, University of Turku, FI-20014 Turun Yliopisto, Finland\label{inst97}\and
Aalto University Metsähovi Radio Observatory, Metsähovintie 114, FI-02540 Kylmälä, Finland\label{inst98}\and
Institute for Astrophysical Research, Boston University, 725 Commonwealth Ave., Boston, MA 02215, USA\label{inst99}\and
Korea National University of Science and Technology, Gajeong-ro 217, Yuseong-gu, Daejeon 34113, Republic of Korea\label{inst100}\and
National Institute of Technology, Ichinoseki College, Takanashi, Hagisho, Ichinoseki, Iwate, 021-8511, Japan\label{inst101}\and
Joint Institute for VLBI ERIC (JIVE), Oude Hoogeveensedijk 4, 7991 PD Dwingeloo, The Netherlands\label{inst102}\and
CSIRO, Space and Astronomy, PO Box 76, Epping, NSW 1710, Australia\label{inst103}\and
Department of Physics, Ulsan National Institute of Science and Technology (UNIST), Ulsan 44919, Republic of Korea\label{inst104}\and
Department of Physics, Korea Advanced Institute of Science and Technology (KAIST), 291 Daehak-ro, Yuseong-gu, Daejeon 34141, Republic of Korea\label{inst105}\and
Kogakuin University of Technology \& Engineering, Academic Support Center, 2665-1 Nakano, Hachioji, Tokyo 192-0015, Japan\label{inst106}\and
Max-Planck-Institut für Astrophysik, Karl-Schwarzschild-Str. 1, 85748 Garching, Germany\label{inst107}\and
Graduate School of Science and Technology, Niigata University, 8050 Ikarashi 2-no-cho, Nishi-ku, Niigata 950-2181, Japan\label{inst108}\and
Anton Pannekoek Institute for Astronomy, University of Amsterdam, Science Park 904, 1098 XH, Amsterdam, The Netherlands\label{inst109}\and
Physics Department, National Sun Yat-Sen University, No. 70, Lien-Hai Road, Kaosiung City 80424, Taiwan, R.O.C.\label{inst110}\and
David A. Dunlap Department of Astronomy \& Astrophysics, University of Toronto, 50 St. George St, M5S 3H4, ON, Canada\label{inst111}\and
Dunlap Institute for Astronomy and Astrophysics, University of Toronto, 50 St. George Street, Toronto, ON M5S 3H4, Canada\label{inst112}\and
School of Astronomy and Space Science, Nanjing University, Nanjing 210023, People's Republic of China\label{inst113}\and
Key Laboratory of Modern Astronomy and Astrophysics, Nanjing University, Nanjing 210023, People's Republic of China\label{inst114}\and
INAF-Istituto di Radioastronomia, Via P. Gobetti 101, I-40129 Bologna, Italy\label{inst115}\and
Common Crawl Foundation, 9663 Santa Monica Blvd. 425, Beverly Hills, CA 90210 USA\label{inst116}\and
Instituto de Física, Pontificia Universidad Católica de Valparaíso, Casilla 4059, Valparaíso, Chile\label{inst117}\and
Key Laboratory of Radio Astronomy and Technology,  Shanghai Astronomical Observatory, CAS, 80 Nandan Road, Shanghai 200030, People’s Republic of China\label{inst118}\and
INAF-Istituto di Radioastronomia \& Italian ALMA Regional Centre, Via P. Gobetti 101, I-40129 Bologna, Italy\label{inst119}\and
Department of Physics, National Taiwan University, No. 1, Sec. 4, Roosevelt Rd., Taipei 106216, Taiwan, R.O.C\label{inst120}\and
Department of Physics and Astronomy, University of Mississippi, Mississippi 38677, USA\label{inst121}\and
Yunnan Observatories, Chinese Academy of Sciences, 650011 Kunming, Yunnan Province, People's Republic of China\label{inst122}\and
Center for Astronomical Mega-Science, Chinese Academy of Sciences, 20A Datun Road, Chaoyang District, Beijing, 100012, People's Republic of China\label{inst123}\and
Key Laboratory for the Structure and Evolution of Celestial Objects, Chinese Academy of Sciences, 650011 Kunming, People's Republic of China\label{inst124}\and
Gravitation and Astroparticle Physics Amsterdam (GRAPPA) Institute, University of Amsterdam, Science Park 904, 1098 XH Amsterdam, The Netherlands\label{inst125}\and
Institute of Astronomy, University of Cambridge, Madingley Road, Cambridge CB3 0HA, United Kingdom\label{inst126}\and
Deceased\label{inst127}\and
Joint ALMA Observatory, Alonso de C\'ordova 3107, Vitacura 763-0355, Santiago, Chile\label{inst128}\and
European Southern Observatory, Alonso de C\'ordova 3107, Vitacura, Casilla 19001, Santiago, Chile\label{inst129}\and
School of Physics and Astronomy, Shanghai Jiao Tong University, 800 Dongchuan Road, Shanghai, 200240, People’s Republic of China\label{inst130}\and
Soft Computing, Image Processing and Aggregation Research Group (SCOPIA) \& Modelling and Imaging Radio Astronomical Data (MIRADA), Dept. of Mathematics and Computer Science, University of the Balearic Islands, Ctra. Valldemossa, Km 7.5, Palma 07122, Spain\label{inst131}\and
Artificial Intelligence Research Institute of the Balearic Islands (IAIB), Palma 07122, Spain\label{inst132}\and
Health Research Institute of the Balearic Islands (IdISBa), 07010 Palma de Mallorca, Spain\label{inst133}\and
Institut de Radioastronomie Millimétrique (IRAM), Avenida Divina Pastora 7, Local 20, E-18012, Granada, Spain\label{inst134}\and
National Institute of Technology, Hachinohe College, 16-1 Uwanotai, Tamonoki, Hachinohe City, Aomori 039-1192, Japan\label{inst135}\and
SKA Observatory, Jodrell Bank, Lower Withington, Macclesfield, SK11 9FT, UK\label{inst136}\and
Department of Physics, Villanova University, 800 Lancaster Avenue, Villanova, PA 19085, USA\label{inst137}\and
Cavendish Astrophysics, University of Cambridge, Madingley Road, Cambridge CB3 0HA, UK\label{inst138}\and
Kavli Institute for Cosmology, University of Cambridge, Madingley Road, Cambridge CB3 0HA, UK\label{inst139}\and
Physics Department, Washington University, CB 1105, St. Louis, MO 63130, USA\label{inst140}\and
Departamento de Matemática da Universidade de Aveiro and Centre for Research and Development in Mathematics and Applications (CIDMA), Campus de Santiago, 3810-193 Aveiro, Portugal\label{inst141}\and
School of Physics, Georgia Institute of Technology, 837 State St NW, Atlanta, GA 30332, USA\label{inst142}\and
School of Space Research, Kyung Hee University, 1732, Deogyeong-daero, Giheung-gu, Yongin-si, Gyeonggi-do 17104, Republic of Korea\label{inst143}\and
G-LAMP NEXUS Institute, Kyung Hee University, Yongin, 17104, Republic of Korea\label{inst144}\and
Canadian Institute for Advanced Research, 180 Dundas St West, Toronto, ON M5G 1Z8, Canada\label{inst145}\and
Dipartimento di Fisica, Università di Trieste, I-34127 Trieste, Italy\label{inst146}\and
INFN Sez. di Trieste, I-34127 Trieste, Italy\label{inst147}\and
Department of Physics, National Taiwan Normal University, No. 88, Sec. 4, Tingzhou Rd., Taipei 116, Taiwan, R.O.C.\label{inst148}\and
Center of Astronomy and Gravitation, National Taiwan Normal University, No. 88, Sec. 4, Tingzhou Road, Taipei 116, Taiwan, R.O.C.\label{inst149}\and
Signal Processing Research Centre, Tampere University, FI-33720 Tampere, Finland\label{inst150}\and
Department of Mathematics, New Uzbekistan University, Tashkent 100007, Uzbekistan\label{inst151}\and
School of Mathematics, Trinity College, Dublin 2, Ireland\label{inst152}\and
Julius-Maximilians-Universität Würzburg, Fakultät für Physik und Astronomie, Institut für Theoretische Physik und Astrophysik, Lehrstuhl für Astronomie, Emil-Fischer-Str. 31, D-97074 Würzburg, Germany\label{inst153}\and
Department of Physics, University of Toronto, 60 St. George Street, Toronto, ON M5S 1A7, Canada\label{inst154}\and
Department of Physics, Tokyo Institute of Technology, 2-12-1 Ookayama, Meguro-ku, Tokyo 152-8551, Japan\label{inst155}\and
Hiroshima Astrophysical Science Center, Hiroshima University, 1-3-1 Kagamiyama, Higashi-Hiroshima, Hiroshima 739-8526, Japan\label{inst156}\and
Aalto University Department of Electronics and Nanoengineering, PL 15500, FI-00076 Aalto, Finland\label{inst157}\and
Institut de Radioastronomie Millimétrique (IRAM), 300 rue de la Piscine, F-38406 Saint Martin d'Hères, France\label{inst158}\and
Jeremiah Horrocks Institute, University of Lancashire, Preston PR1 2HE, UK\label{inst159}\and
National Biomedical Imaging Center, Peking University, Beijing 100871, People’s Republic of China\label{inst160}\and
College of Future Technology, Peking University, Beijing 100871, People’s Republic of China\label{inst161}\and
Department of Physics and Astronomy, University of Lethbridge, Lethbridge, Alberta T1K 3M4, Canada\label{inst162}\and
Frontier Research Institute for Interdisciplinary Sciences, Tohoku University, Sendai 980-8578, Japan\label{inst163}\and
Astronomical Institute, Tohoku University, Sendai 980-8578, Japan\label{inst164}\and
Department of Physics and Astronomy, Seoul National University, Gwanak-gu, Seoul 08826, Republic of Korea\label{inst165}\and
SNU Astronomy Research Center, Seoul National University, Gwanak-gu, Seoul 08826, Republic of Korea\label{inst166}\and
ASTRON, Oude Hoogeveensedijk 4, 7991 PD Dwingeloo, The Netherlands\label{inst167}\and
Centre for Mathematical Plasma Astrophysics, Department of Mathematics, KU Leuven, Celestijnenlaan 200B, B-3001 Leuven, Belgium\label{inst168}\and
Physics Department, Brandeis University, 415 South Street, Waltham, MA 02453, USA\label{inst169}\and
Tuorla Observatory, Department of Physics and Astronomy, University of Turku, FI-20014 Turun Yliopisto, Finland\label{inst170}\and
Excellence Fellow at Radboud University, Nijmegen, The Netherlands\label{inst171}\and
School of Natural Sciences, Institute for Advanced Study, 1 Einstein Drive, Princeton, NJ 08540, USA\label{inst172}\and
School of Physics, Huazhong University of Science and Technology, Wuhan, Hubei, 430074, People's Republic of China\label{inst173}\and
Mullard Space Science Laboratory, University College London, Holmbury St. Mary, Dorking, Surrey, RH5 6NT, UK\label{inst174}\and
Center for Astronomy and Astrophysics and Department of Physics, Fudan University, Shanghai 200438, People's Republic of China\label{inst175}\and
Astronomy Department, University of Science and Technology of China, Hefei 230026, People's Republic of China\label{inst176}\and
Department of Physics and Astronomy, Michigan State University, 567 Wilson Rd, East Lansing, MI 48824, USA\label{inst177}
}

\authorrunning{H. Mueller, S.-D. von Fellenberg et al.}
\abstract{The 2021 Event Horizon Telescope observations resolve the innermost jet region of the blazar 3C~279 with unprecedented detail. The reconstructed images consistently reveal a compact core elongated nearly orthogonal to the large-scale jet axis. This rarely observed morphology recurs across multiple epochs and from 22-230\,GHz and is therefore intrinsic rather than an imaging artifact. Geometric model fitting identifies several components with apparent speeds up to $\beta_{\rm app} \sim 10$, requiring bulk Lorentz factors $\Gamma \gtrsim 10.3$ and constraining viewing angles to extremely small values ($\lesssim 1^\circ$). Rest-frame brightness temperatures are systematically low ($T_B \sim 10^9$--$10^{10}\,\mathrm{K}$), consistent with optically thin emission at 230\,GHz. These results suggest that the jet bends toward the observer on sub-parsec scales, producing strong relativistic beaming. Possible drivers of the observed jet bending and temporal evolution include the jet's interaction with the interstellar medium, kink or Kelvin--Helmholtz instabilities, magnetic reconnection near the horizon, or binary-induced precession. However, the current temporal coverage of VLBI data remains insufficient to distinguish between these mechanisms. Continued multifrequency VLBI monitoring will be essential to constraining the dynamics and geometry of the jet base in 3C~279.}
\bigskip
\keywords{accretion: accretion disks, black hole physics, gravitation, relativistic process, galaxies: jets}

\date {Received  / Accepted}

\maketitle
\nolinenumbers
%
\section{Introduction}
\label{sec:intro}

Active galactic nuclei (AGNs) host some of the most powerful and long-lived outflows in the Universe. These include relativistic jets, which can transport energy and momentum from the immediate vicinity of a supermassive black hole to scales far beyond their host galaxies. Despite decades of observational and theoretical effort, fundamental questions remain regarding how jets are launched, accelerated, collimated, and stabilized, as well as how they evolve dynamically and how their radiative properties relate to the accretion flow and black hole spin \citep{Blandford2019}.

Due to their high luminosity, blazars, whose jets are oriented at small angles to the line of sight, are particularly valuable targets for millimeter very long baseline interferometry (mm-VLBI). In the millimeter regime, observations of AGNs typically probe optically thin regions that are inaccessible at longer wavelengths due to synchrotron self-absorption. Thereby, mm-VLBI observations allow direct access to the jet base on sub-parsec and even horizon-scales \citep[e.g.,][]{Lu2023, Kim2025, Saurabh2025}. With angular resolutions of tens of microarcseconds, global mm-VLBI arrays such as the Event Horizon Telescope (EHT) enable detailed studies of jet morphology, variability, and magnetic field structure for AGN sources at the spatial scales where jet launching and initial jet collimation occur \citep{Kim2020, Janssen2021, Issaoun2022, Jorstad2023, Paraschos2024, Saurabh2025, Traianou2025,Gomez2026}.

3C~279 is an archetypal blazar located at a distance of approximately $\sim 16.5~\mathrm{Mpc}$  \citep[$z\approx0.536$, e.g,][]{sloan72009,Bird2010,Lister2018} corresponding to a projected distance scale of $\sim 11.4~\mathrm{pc/mas}$ and a deprojected scale of $\sim340~\mathrm{pc/mas}$ for an assumed viewing angle of $1^\circ$ \citep[based on 43 GHz monitoring][]{Jorstad2017}. Its relativistic jet is powered by a supermassive black hole with an estimated mass of  
$\sim 8 \times 10^{8}~\mathrm{M_\odot}$ \citep[e.g.,][]{Nilsson2009}. Our observations thus constrain jet-launching well within the Bondi-accretion region of 3C~279 ($800 \times r_{\rm{g}} \simeq ~ 10 \mu\mathrm{as}$), scales which are accessible to modern, large-scale simulations of accreting black holes \citep[e.g.,][]{Tchekhovskoy2016kinky,Lalakos2024}. The source shows an approximate ``blue-bump'' luminosity of $2\times10^{45}~\mathrm{erg~s^{-1}}$ \citep[e.g.,][]{Pian1999}, yielding an Eddington accretion rate of $\approx0.02~\mathrm{\dot{m}_{\rm{Edd.}}}$. This indicates that the source accretes radiatively efficiently in the thin or truncated thin disk regime, consistent with its appearance as a flat spectrum radio quasar \citep[FSRQ; e.g.,][]{Urry1995}. 

Owing to its brightness, 3C~279 has long served as a prime target for VLBI and is frequently observed as a calibrator source. The jet has been systematically monitored by programs such as the Boston University Blazar Monitoring project \citep{Jorstad2017} and MOJAVE \citep{Lister2018}, which, together with complementary studies across the full electromagnetic spectrum, have established key physical parameters. VLBI observations indicate a small viewing angle  \citep{Jorstad2017}, a VLBI core (i.e., the brightest VLBI component, not necessarily the physical jet origin) with an observed brightness temperature exceeding $10^{12}\,\mathrm{K}$ \citep[$\log_{10}(T_{B})=12.8^{+0.1}_{-0.2}$, ][]{Homan2021}, and strong polarization, with linear fractions above 15\% and circular fractions above 1\% at frequencies below 43\,GHz \citep{Homan1999, Homan2001, Homan2009, Vitrishchak2008}. High circular polarization ($\sim 3\%$) has also been confirmed at 86\,GHz using single-dish monitoring within the POLAMI project performed at the IRAM 30m telescope \citep{Thum2018}. Estimates of the bulk Lorentz factor place the jet speed at larger scales in the range $\Gamma \sim 10 \text{–} 70$ \citep{Homan2003, Bloom2013, Homan2015, Jorstad2017, Lister2019, Lister2021,Homan2021,Weaver2022}, while $\gamma$-ray observations require $\Gamma > 35$, with favored values around $\Gamma \sim 50$ to account for the overall spectral energy distribution (SED) shape of the source \citep{Ackermann2016}.

High-resolution mm-VLBI and space-based centimeter VLBI have provided direct constraints on the innermost jet structure. The EHT campaigns, beginning with proto-EHT data in 2011 \citep{Lu2013}, followed by the 2017 observations \citep{Kim2020} and subsequent 2018 data (in prep.), have resolved the VLBI core at angular scales down to $\sim 20\,\mu\mathrm{as}.$ These data reveal an elongated morphology that is difficult to reconcile with simple jet models, suggesting possible precession or the presence of a compact acceleration zone within the inner $20\,\mu\mathrm{as}$. The EHT 2017 observations were accompanied by one of the most extensive quasi-simultaneous multiwavelength campaigns of this source \citep{Principe2026}. From April 5–11, 2017, EHT measurements revealed a flux increase in the innermost VLBI core (i.e., the kinematic origin), while longer-wavelength radio data over a broader time range showed enhanced core flux and polarization, coincident with the ejection of a superluminal knot. This followed a record UV–optical flare with strong polarization variability in late March, accompanied by intense $\gamma$-ray flaring activity.

Space-VLBI observations with RadioAstron extend this picture to centimeter-wavelengths. Filamentary structures were detected in 2014 \citep{Fuentes2023} roughly $100\,\mu\mathrm{as}$ downstream in the jet and show an elongated core-structure. Nonetheless, this structure was not recovered with RadioAstron observation taken in 2018, likely due to sensitivity limitations and poor uv coverage \citep{Toscano2025}. The EHT and RadioAstron results provide a multi-scale view: within the inner $\sim 20\,\mu\mathrm{as}$, the VLBI core is elongated and complex, and the jet most likely accelerates rapidly close to the kinematic origin. Between $20-100\,\mu\mathrm{as}$, a strongly bent jet is observed, leading to components that align with downstream jet emission, consistent with the cm-VLBI observations of the milliarcsecond scale jet.

At scales of $100\,\mu\mathrm{as}-1\,\mathrm{mas},$ the jet transitions into a cylindrical channel observed across a wide frequency range, likely formed by external pressure confinement \citep{Homan2003, Kellermann2004, Jorstad2004,  Homan2009b, Jorstad2017, Lister2018, Lister2021}. The reported bulk-flow speed in this region reaches $\sim 20c.$ At milliarcsecond scales, the jet shows varying position angles, potentially described by instabilities forming downstream in the jet.

\begin{figure*}
    \centering
    \includegraphics[width=\textwidth]{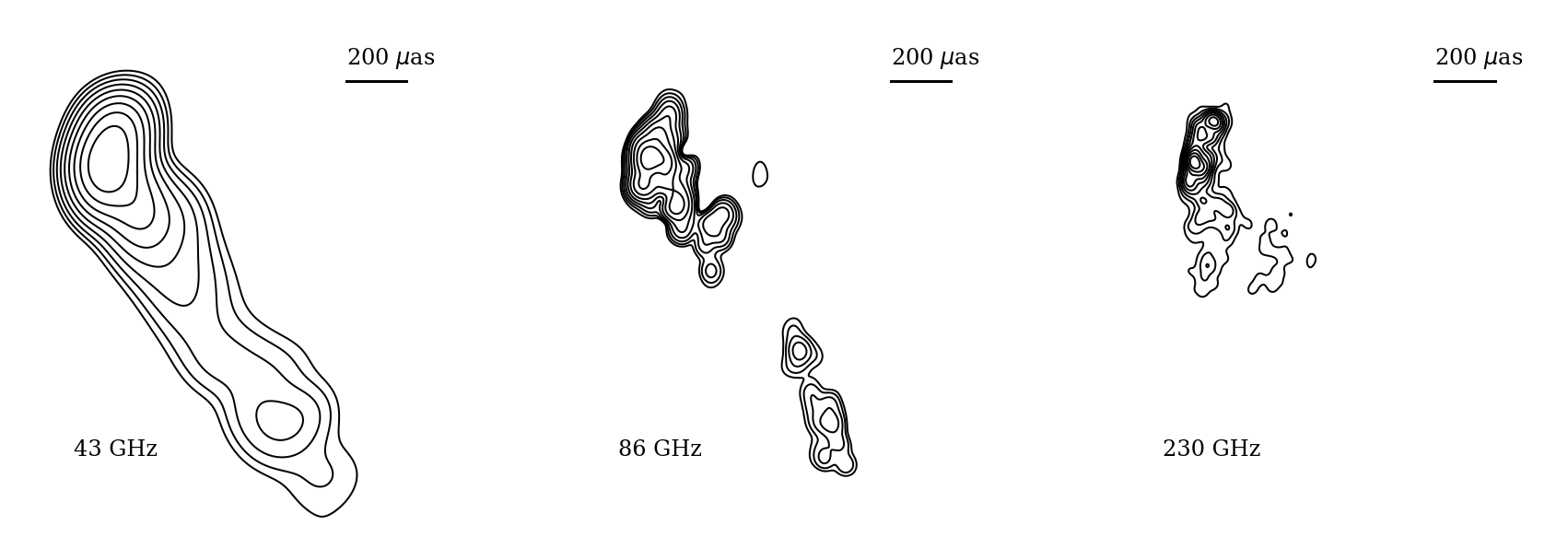}
    \caption{3C~279 at three different frequencies from quasi-simultaneous observations in April 2021. Left panel: VLBA observations of the BEAM-ME program at 43\,GHz at April 5 (obtained by \texttt{kine}). Middle panel: Preliminary GMVA observations at 86\,GHz obtained at April 22 (obtained by \texttt{Difmap}). Right panel: EHT observations at 230\,GHz on April 13 (obtained by \texttt{Comrade}). The contour levels increase by a factor of 2 from $0.19\%$ to $50\%$ of the respective peak flux. The peak total intensity values are 6.7 $\mathrm{Jy/{\left(116\,\mu\mathrm{as}\right)^2}}$ at 43 GHz, 7.2 $\mathrm{Jy/{\left(66\,\mu\mathrm{as}\right)^2}}$ at 86 GHz, and 2.5 $\mathrm{Jy/{\left(25\,\mu\mathrm{as}\right)^2}}$ at 230 GHz.}
    \label{fig:multifrequency}
\end{figure*}

In this work, we present new observations taken by the EHT at 230 GHz in the 2021 observation campaign. EHT achieves unprecedented resolution of the inner nuclear structure. Combined with earlier EHT epochs, and multifrequency data, we measured the changes in the inner nuclear structure, which helps identify its nature. We show our images in Fig. \ref{fig:multifrequency}. Throughout this paper we adopt a cosmology with $H_0 =67.7\,\mathrm{km}\mathrm{s}^{-1}\mathrm{Mpc}^{-1}$, $\Omega_m=0.307$, and $\Omega_\Lambda=0.693$ \citep{Planck2016}.

\section{Observations and data}

The analysis presented in this paper is based on EHT observations obtained in 2021 at around 230\,GHz. Three observing dates were considered: April 13, April 17, and April 18. The April 13 and April 18 tracks correspond to the same datasets used in the EHT M87 2021 campaign \citep{M87_2021}, while the April 17 track has not been analyzed in previous publications. In all tracks, 3C~279 served as one of the primary calibration sources. The science targets were M87 on April 13 and April 18, and Sgr~A* on April 17.
The 2021 campaign represents the largest EHT array configuration published to date. In addition to the established stations, two new sites participated: the Kitt Peak 12\,m telescope (KP) and the NOEMA (NN) interferometer in the French Alps \citep[e.g.,][]{Pietu2025NoemaPhasing}. These additions provided crucial intermediate baselines that significantly improved the $(u,v)$ coverage; see \autoref{fig:uvcov}. In particular, they provide intermediate-scale baselines which enabled constraints on the angular scales of 
$200~\mu\mathrm{as} \quad \text{and} \quad 2000~\mu\mathrm{as},$ \citep{M87_2021,Saurabh2025, Georgiev2025}. These constraints are essential for probing both compact and extended structures in the target sources. 3C~279 was observed in four bands centered at 213.1, 215.1, 227.1, and 229.1 GHz, each with a bandwidth of 1875 MHz (hereafter referred to as bands 1--4). In this manuscript, we present the results obtained from the upper two bands (bands 3 and 4).
All data were fringe-fitted and calibrated using the \texttt{rPICARD} calibration pipeline within CASA \citep{Janssen2019}. The calibration procedures follow the methodology described in detail in \citet{M87_2021, vonFellenberg2025}, including amplitude and phase calibration, band-pass corrections, and polarization leakage mitigation. The data were released in \cite{M87_2021} and are available in the ALMA archive (Observing Program 2019.1.01797.V).

\section{Imaging and modeling}
The calibrated visibilities from the 2021 EHT campaign were imaged and modeled using a suite of independent algorithms. We employed three distinct codes: \texttt{Comrade} \citep{Tiede2022}, \texttt{DoG-HiT} \citep{Mueller2022}, and \texttt{kine} (in prep.). Each code independently produced one image per epoch and per band, resulting in a set of reconstructions for April 13, April 17, and April 18, respectively (18 images in total).

The imaging pipelines differ in their regularization strategies, optimization schemes, and parameterizations of source structures. \texttt{Comrade} \citep{Tiede2025} implements a Bayesian inference framework with hierarchical modeling of visibility amplitudes and closure phases, while \texttt{DoG-HiT} \citep{Mueller2022, Mueller2023a, Mueller2023b} employs a gradient-based optimization with a wavelet-based sparsity prior. \texttt{kine} is a recently proposed imaging algorithm based on unsupervised machine learning avoiding explicit regularizers. The diversity of algorithms ensures that the reconstructed features are not artifacts of a single method. Moreover, these newer algorithms are usually more automatized and robust against user calibration choices, making them well suited for EHT data. \texttt{DoG-HiT} and \texttt{Comrade} have been already utilized in the analysis of M87 from the same observations \citep{M87_2021}. In Sect. \ref{sec:imaging}, we discuss validating our reconstructions with more traditional VLBI imaging approaches (using \texttt{Difmap}).
For each epoch, the imaging teams submitted a single representative reconstruction per code. These images were compared across pipelines to assess the robustness of morphological features. Despite the inclusion of the Kitt Peak and NOEMA stations in 2021, which provide constraints on the extended emission at intermediate scales ($200~\mu\mathrm{as} \quad \text{and} \quad 2000~\mu\mathrm{as}$), the EHT data exhibits a u-v coverage that is too sparse to robustly reconstruct the faint, large-scale structure detected at longer wavelengths. 

In addition, we performed model fitting using \texttt{Difmap}, providing parametric fits to the visibilities using elliptical Gaussian components, where the choice and location of the components were motivated by the images. The quoted model fit uncertainties are based on the statistical fit uncertainty derived by \texttt{Difmap}.

\section{Results}

\subsection{Imaging} \label{sec:imaging}

We chose the images obtained by the \texttt{comrade} software as our fiducial set, since they consist of the only imaging approach that also calculates a relative uncertainty. We show the relative uncertainty derived by the \texttt{comrade} posterior in Appendix \ref{app:additional_reconstruction}. We utilized imaging results obtained by independent pipelines as validation tests. As an example, we show the fiducial image obtained from band 3 (226.1 GHz) on April 13 in Fig. \ref{fig:e21e13_b3}, together with the images obtained by alternative imaging and modeling approaches. 
Additional reconstructions from different dates and band 4 are shown in Appendix \ref{app:additional_reconstruction}. Bands 3 and 4 were imaged independently from each other, serving as an additional consistency check. The source structure is consistent across most approaches and bands, including multiple features resolved, their orientation, spacing and brightness ratios. We note, however, that the traditional CLEAN algorithm implemented in \texttt{Difmap} finds two degenerate solutions (shown in Fig. \ref{fig:clean_vs_comrade}), which result from different self-calibration strategies. Only one solution is consistent with the remainder of the imaging codes. CLEAN is more dependent on the iterative, manual user-based choices regarding flagging, masking, and affecting the self-calibration. Since this cleaning, masking, and flagging in \texttt{Difmap} is an iterative, manual procedure, it is challenging to identify a single difference which caused the degeneracy. The result shown in the right panel in Fig. \ref{fig:clean_vs_comrade} (hereafter \texttt{Difmap 2}), however, differs from the solution shown in the middle panel (\texttt{Difmap 1}) due to the addition of a large-scale Gaussian in \texttt{Difmap}, aggressive flagging of the baselines including the GLT, and the requirement for larger amplitude gain corrections than in \texttt{Difmap 1}.

\begin{figure}
    \centering
    \includegraphics[width=\linewidth]{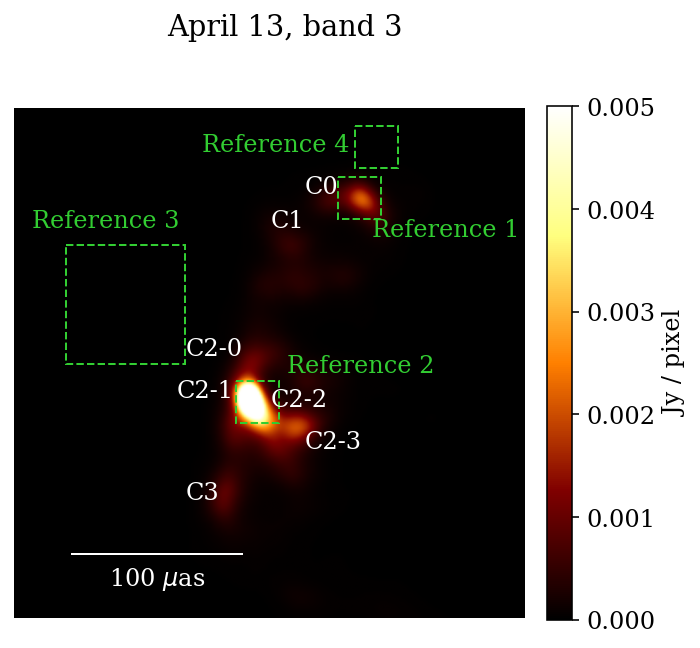}
    \caption{Naming convention of the \texttt{Difmap} model components, based on the Comrade image from April 13, band 3. The C2 structure is modeled by four components. We also sketch three possible reference positions for the kinematical analysis, which are discussed in Sect. \ref{sec:core_identification}. The pixel size is $1\,\mu\mathrm{as}^2$.}
    \label{fig:annotation}
\end{figure}

\begin{figure}
    \centering
    \includegraphics[width=\linewidth]{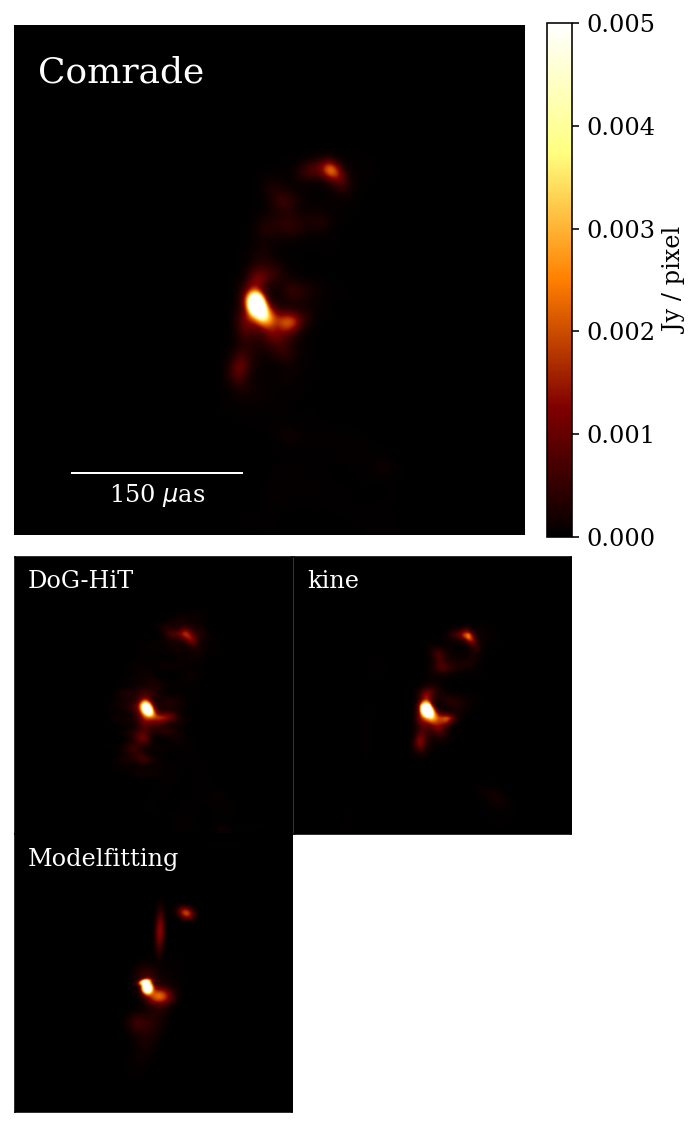}
    \caption{Fiducial image of the source in linear scale for the first observing epoch (April 13) at 226 GHz taken with EHT. The large image shows the \texttt{Comrade} reconstruction, while the lower figures show qualitatively similar reconstructions based on \texttt{DoG-HiT}, \texttt{kine}, and \texttt{Difmap} model fitting. April 13 is chosen as a representative example. Nevertheless, the imaging results are consistent throughout the different days.}
    \label{fig:e21e13_b3}
\end{figure}

\begin{figure}
    \centering
    \includegraphics[width=\linewidth]{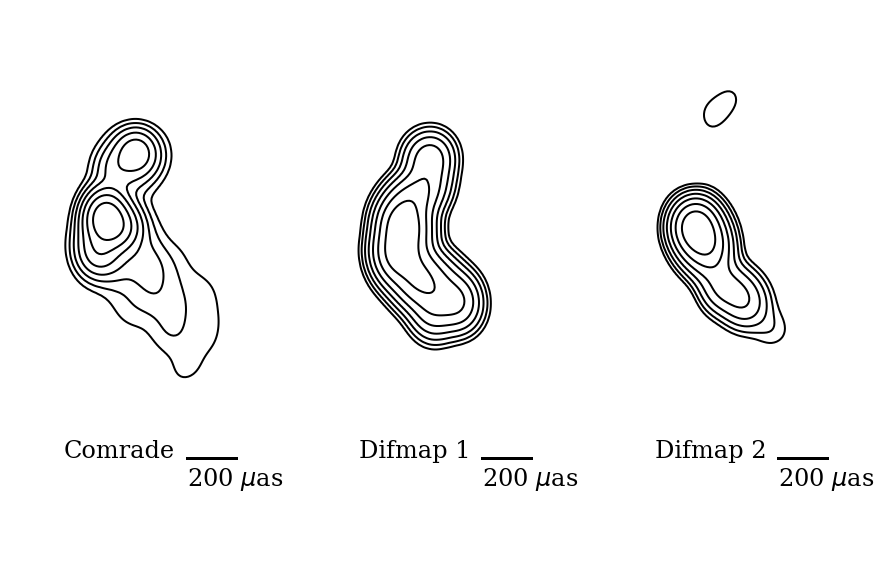}
    \caption{Reconstruction results in band 3 on April 13 with \texttt{Comrade} and two different \texttt{Difmap} solutions with comparable $\chi^2$. The images were blurred to the same resolution and show equal contour levels. The contour levels increase by a factor of 2 from $0.19\%$ to $50\%$ of the respective peak total intensity. The respective peak fluxes are $3.6$, $1.7$, and $3.7$ $\mathrm{Jy/{\left(25\,\mu\mathrm{as}\right)^2}}$.}
    \label{fig:clean_vs_comrade}
\end{figure}

The source structure can be broadly classified into four groups of components, as illustrated in Fig.~\ref{fig:annotation}. At the center, we identify several bright features (labeled C2-x), which exhibit distinct "hook"-like bends. Toward the north, a fainter component (C0) is visible, with indications of faint emission bridging it to the central region (labeled C1). To the south, another weaker feature (C3) can be distinguished. The total flux density of the source amounts to $9.96 \pm 0.02\,\mathrm{Jy}$.

\begin{figure*}
    \centering
    \includegraphics[width=\textwidth]{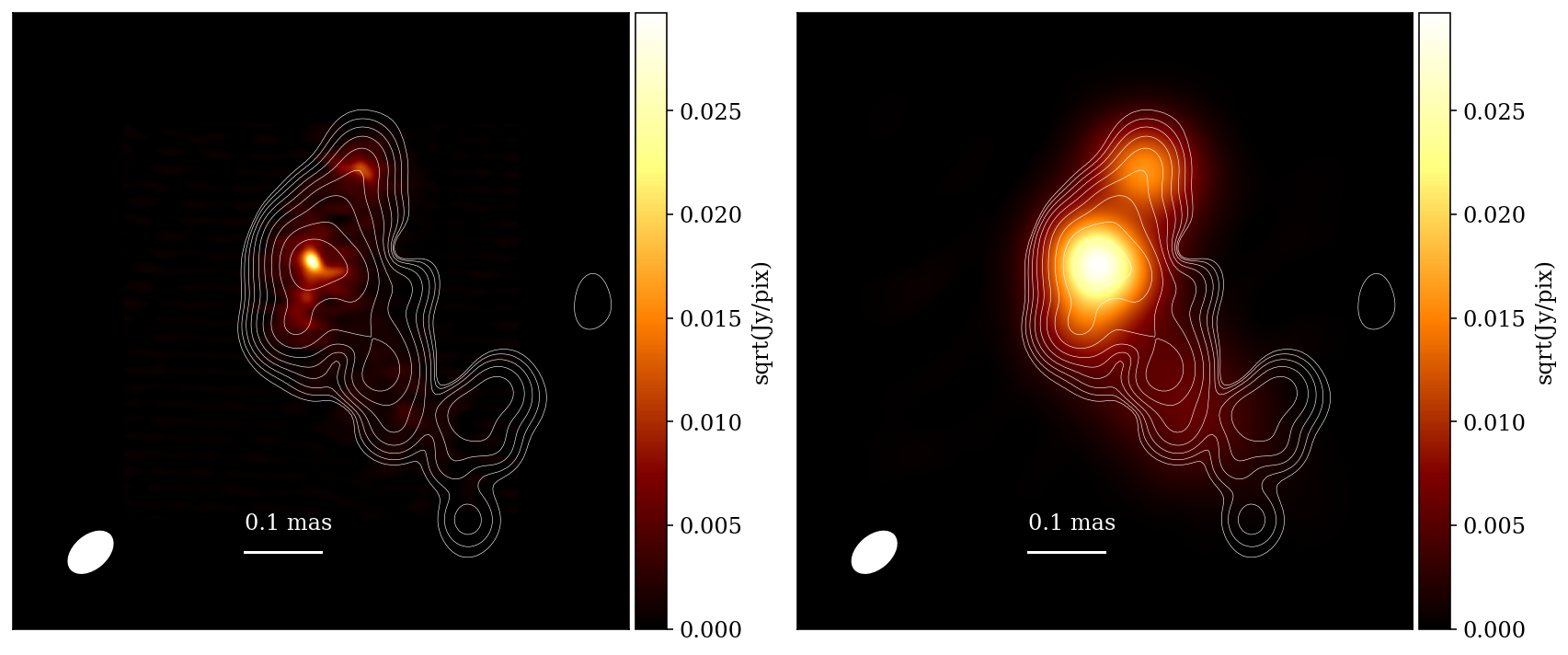}
    \caption{230\,GHz image of 3C~279 observed in 2021 compared to the structure observed by the GMVA in 2021 (in prep.). The EHT image is shown in color, overlaid with 86\,GHz GMVA contours. Left: Native EHT resolution. Right: EHT image convolved with a circular $66\,\mu\mathrm{as}$ beam, maximizing the cross-correlation with the GMVA image (GMVA beam: $72 \times 47 \mu\mathrm{as}$. The contour levels increase by a factor of 2 from $0.19\%$ to $50\%$ of the respective peak total intensity.}
    \label{fig:fiducial}
\end{figure*}

To further verify this source structure, we compared the EHT images with the quasi-simultaneous GMVA and VLBA observations taken at 43\,GHz and 86\,GHz, as shown in Fig. \ref{fig:multifrequency}. The image morphology is consistent across both frequencies, including the central structure (C2-x), northern (C0) component, southern (C3) component, and diffuse emission toward the southeast. An overlay of the GMVA (including the phased-ALMA and GLT, in prep.) and EHT image is shown in Fig. \ref{fig:fiducial}. The two images were aligned by maximizing their correlation. For a detailed analysis of the image at 86 GHz and the spectral index, we refer to Kim et al., in prep.
The VLBA data stem from the Boston University blazar monitoring program (BEAM-ME) dataset\footnote{\url{https://www.bu.edu/blazars/VLBA_GLAST/3c279.html}} (left panel of Fig. \ref{fig:multifrequency}), imaged using the super-resolving dynamic imaging code \texttt{kine}. This code achieves higher resolution and higher fidelity by combining data from multiple epochs rather than obtaining individual snapshot images. Multiple images confirm the source morphology at 43\,GHz (see App.~\ref{app:kine iamges} and Fig. \ref{fig:43GHz}). 

\begin{figure}
    \centering
    \includegraphics[width=\linewidth]{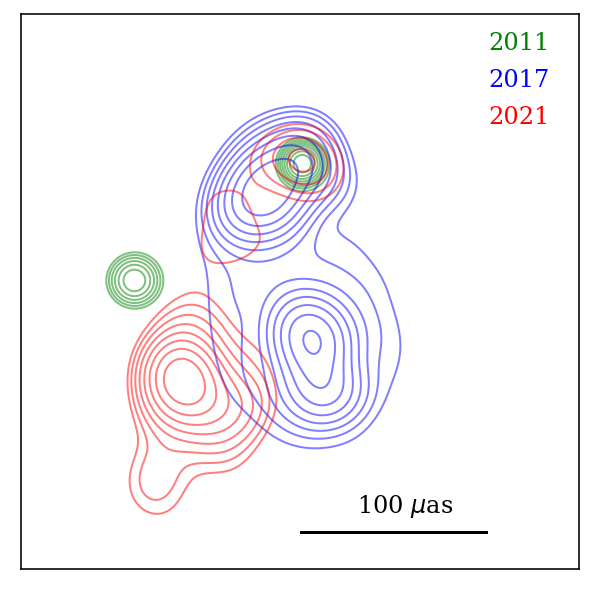}
    \caption{1.3 mm structure of 3C~279 obtained at three different years with the 2011 EHT data presented in \citet{Lu2013}, the 2017 EHT data in \citet{Kim2020}, and the 2021 EHT data reported in this manuscript. All these studies report the northernmost component as the most likely kinematic origin. Therefore, the alignment in this figure was performed on the northernmost jet component, assumed to be the stationary VLBI core. The contour levels increase by a factor of 2 from $0.19\%$ to $50\%$ of the respective peak total intensity.}
    \label{fig:multiyear}
\end{figure}

Finally, Fig.~\ref{fig:multiyear} compares the 2021 source image to earlier EHT epochs. The 2011 proto-EHT data model-fitting results \citep{Lu2013} are shown in green contours, while the structure recovered from the 2017 EHT observations \citep{Kim2020} are shown in blue. The 2021 data presented in this paper are shown in red. The images are aligned on the kinematic origin identified in previous multiyear studies \citep{Lu2013, Kim2020}, by registering all epochs to the northern feature, which is taken as the kinematic origin and implicitly assumes a net southward jet motion. The position angle of the jet component closest to the kinematic origin changes largely across these three years. The north–south extension seen in 2021 is consistent with the range of structures seen in other epochs.

In summary, the structure identified in the 2021 EHT observations -- particularly the symmetric northern and southern extensions flanking the bright central C2 region (designated C0 and C3) -- is supported by several independent lines of evidence. The morphology consistently appears across multiple imaging algorithms, frequency bands, and observing epochs (see Fig. \ref{fig:images}), including methods relying solely on closure quantities (e.g., \texttt{DoG-HiT}) and parametric model fitting. It is further corroborated by quasi-simultaneous 43\,GHz and 86\,GHz imaging, with the north–south extension persisting over several months in the 43\,GHz data. While it is possible to fit the visibilities without C0 using the \texttt{CLEAN} algorithm (see Fig. \ref{fig:clean_vs_comrade}), this requires more aggressive gain calibration. Taken together, the modest gain corrections required, the algorithmic and temporal consistency, and the multifrequency agreement suggest that both C0 and C3 are plausible intrinsic features of the source morphology.

\subsection{Model fitting}\label{sec:modelfitting}

In Fig. \ref{fig:models}, we show the model-fitting results for every individual dataset, again obtained independently from each other. We note that the components C0, C2-0, C2-1, C2-2, C2-3, and C3 are well constrained and consistent across bands and days. The C1 component is less well constrained due to its faintness. This assessment is supported by an investigation of the relative uncertainties derived by \texttt{Comrade} (see Fig. \ref{fig:comrade_uncertainty}), which indicates that the emission between C0 and C2 is at the detection limit. We therefore ignore this component in our following analysis of the jet's dynamics. We performed the modeling independently for bands 3 and 4 and used the results from the two bands for cross-validation.

\begin{figure*}
    \sidecaption
    \includegraphics[width=12cm]{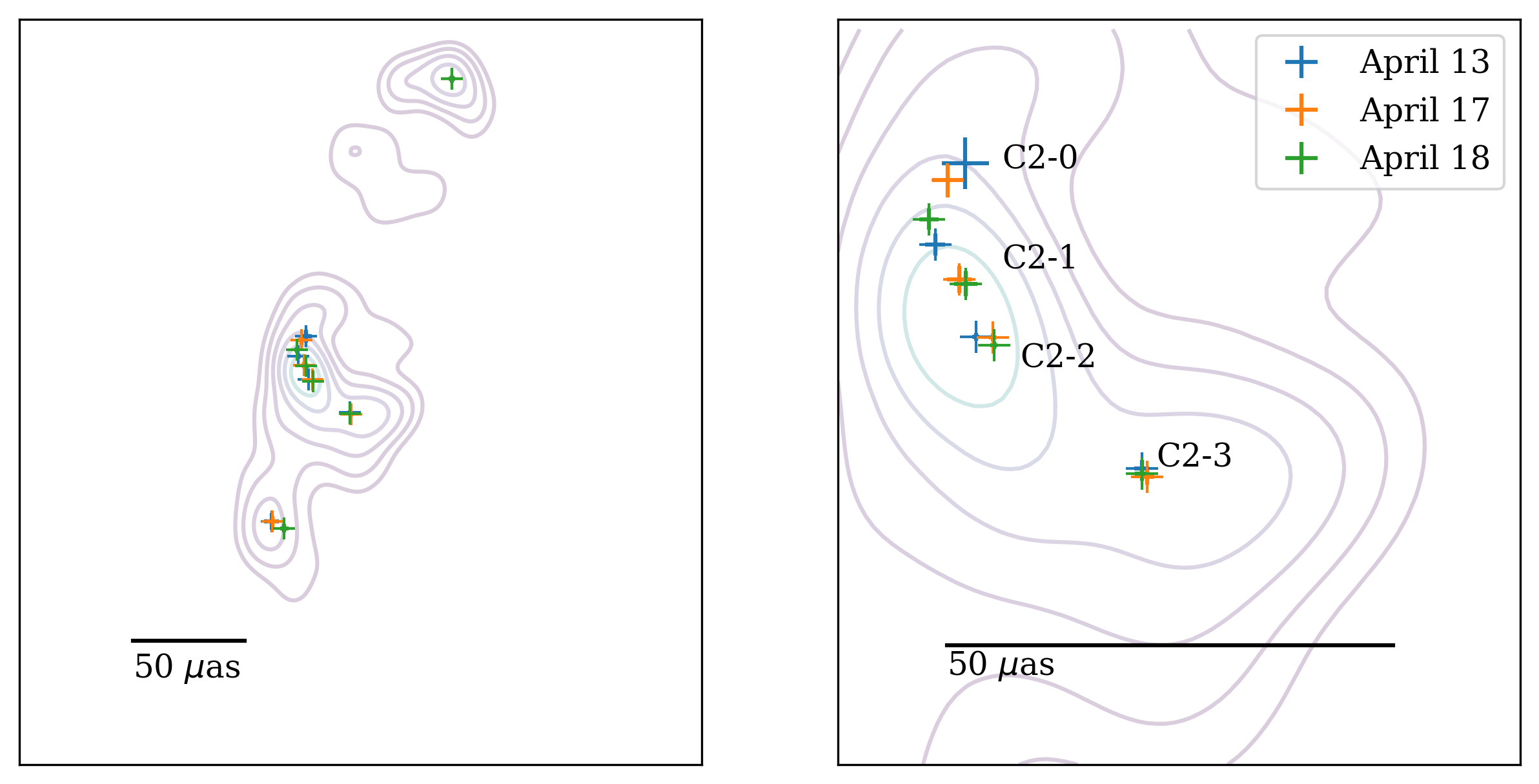}
    \caption{Location of the model-fitted components for band 3. The blue, orange, and green crosses mark the positions at the three different observing epochs. The marker size indicates the model statistical fit uncertainty derived from half the beam size divided by the median signal-to-noise ratio. The contours correspond to the 230\,GHz image reconstructed with \texttt{Comrade}. The two panels display the image at different zoom levels. The contour levels increase by a factor of 2 from $1.56\%$ to $50\%$ of the respective peak total intensity. All maps and components shown were aligned on C0.}
    \label{fig:components}
\end{figure*}

We show the positions of our model components across all three days in Fig. \ref{fig:components} for band 3 (left to right, zoomed-in on the brightest cluster of components). For cross-validation purposes, we compare our independently derived modeling components from bands 3 and 4 in Appendix \ref{app:additional_reconstruction}. We note that the results for both bands match well. The fitted parameters are shown in \autoref{tab:components}. We adapted half the beam size divided by the median signal to noise ratio (S/N) as the positional as well as for the major and minor axes. The error in flux is derived as $\sigma_\mathrm{S}^2 = \sigma_\mathrm{thermal}^2+f_\mathrm{cal}^2 S^2$, where $S$ is the estimated flux of the component, $\sigma_\mathrm{thermal}$ is the thermal noise, and $f_{cal} \sim 0.1$ is the median gain variation. The position angle error is $\sigma_{PA} = \frac{a}{a-b} \frac{1}{2 S/N}$, where $a$ is the major axis and $b$ is the minor axis. Because the measurements do not contain absolute astrometric information, we aligned the images and assumed that the C0 component is the stationary kinematic origin, i.e., it shows no intrinsic motion. Under this assumption, the data show coherent evolution across three days with projected velocities of $1-2\,\mu\mathrm{as}$ per day (corresponding to a deprojected velocity of $\sim 0.34\mathrm{pc/d}$ for $z=0.567$ and $i=1^\circ$ and to apparent velocities of up to $10\,c$). This result is consistent with the projected motion found in the 2017 EHT data at comparable spatial scales \citep{Kim2020}. Similar velocities were observed at lower frequencies and larger spatial scales in the BU-Blazar monitor program \citep{Jorstad2017}. We discuss the dynamics of the jet components in more detail in Sect.~\ref{sec:astrometry}.

\subsection{Identification of kinematic origin} \label{sec:core_identification}

Our imaging analysis of the 2021 EHT data reveals several distinct components. The brightest features (C2) form a central cluster with a western extension aligned to the large-scale jet, while a fainter northern component (C0) is also detected. A key question is which of these corresponds to the true kinematic origin, i.e., the stationary core -- typically used as a proxy for the black hole location -- situated downstream from the jet and core position \citep[e.g.,][]{Plavin2019_gaiaVLBI}. Since no objective criterion exists, the identification must rely on interpretation.

At first glance, the central cluster might appear to be the kinematic origin due to its brightness, but its morphology suggests downstream knots. The northern component, although fainter, is located at a position consistent with the upstream end of a bent jet and could therefore represent the kinematic origin (see kinematic origin 1 in Fig. \ref{fig:annotation}). Multifrequency VLBI comparisons (Fig.~\ref{fig:fiducial}) reveal this northern extension at 86\,GHz and 230\,GHz, supporting its physical reality and suggesting it as a possible core location. Moreover, a reanalysis of the 43 GHz data with the dynamic imaging code \texttt{kine} has revealed that the VLBI core appears elongated in the north-source direction at 43 GHz for multiple months (see Fig. \ref{fig:43GHz}).
We note that similarly misaligned inner structures have been reported in earlier EHT and mm-VLBI and \citep{Lu2013, Kim2020} RadioAstron observations \citep{Fuentes2023, Toscano2025} as well as in quasi-simultaneous 43 and 86\,GHz data, suggesting that this morphology is intrinsic rather than an imaging artifact. Nevertheless, some caution is warranted, as the northern component is not robustly recovered by all imaging methods.
Comparable bent or transverse inner jets where the brightest component does not match the core location have also been observed in other blazars during flaring states \citep{Lobanov2005,Agudo2012,Fromm2013,Hodgson2017,Larionov2020}, pointing to a potentially common mechanism. We also highlight the similarities in the appearance of 3C279 and the compact symmetric object OQ208 \citep{Wu2013}.

An alternative interpretation is that the northern C0 component is itself a jet feature, with the central C2--2 structure marking the true kinematic origin (i.e., reference position 2 in Fig. \ref{fig:annotation}). Theoretical jet--to--counter-jet brightness ratios for 3C~279 exceed $R > 10^{10}$ \citep{Urry1995}, implying that any counter-jet emission would be far below the EHT detection threshold. We therefore exclude the possibility that C0 represents a counter-jet feature. Projection effects, however, cannot be ruled out. A jet component located downstream of C2--x could appear north of it in projection if the flow exhibits filamentary substructures \citep{Fuentes2023}, particularly given the extremely small viewing angle of 3C~279. Still, the observed relative motions between the different components seem to contradict this interpretation. We show in Fig. \ref{fig:components_alternative_alignment} modeling results without C0, with the brightest component C2-2 as the kinematic origin. We note that this identification would imply inward motion for component C2-3, thereby supporting the interpretation of C0 as the kinematic origin. 

A different possibility is that the true core or jet-launching region is not detected at all. In this scenario, all observed components would correspond to emission features propagating downstream in the jet. The kinematic origin would then be associated with the upstream extension of the jet observed with cm-VLBI experiments \citep[e.g., by MOJAVE;][]{Lister2018}, approximately at the sketched reference position 3 in Fig. \ref{fig:annotation}. This interpretation is plausible in light of the high Lorentz factors inferred for the source, which may render the jet-launch region too faint to appear in the relatively low-contrast EHT images. If correct, this would imply that all detected features correspond to optically thin synchrotron emission and should therefore exhibit negative spectral indices on the order of $\sim -1$. This scenario is thus directly testable by comparing the EHT, GMVA, and VLBA quasi-simultaneous data (in prep.).

For the remainder of this manuscript, we assume that the component C0 is the kinematic origin, i.e., the VLBI core which is often interpreted as  the jet base or jet apex. We note that the kinematic origin may coincide with the true location of the AGN central engine at 230 GHz, similar to the interpretation proposed for Messier 87 \citep{Lu2023,M87_2021,Saurabh2025}. Alternatively, it may trace the innermost jet region where the outflow first becomes observable at high radio frequencies while undergoing collimation and acceleration, such that relativistic Doppler boosting renders the emission detectable. In this scenario, the physical jet apex and therefore the true core position would be expected to lie further upstream, potentially near reference position 4 in Fig. 2. A detailed investigation of this possibility, based on spectral index measurements, will be presented in forthcoming work.

\subsection{Brightness temperatures}
The apparent brightness temperature of the C0 and C2 components are found to be $\sim 5 \cdot 10^{10}\,\mathrm{K}$. These values were obtained using
\[
T_B = 1.22 \cdot 10^{12} \frac{F}{\nu^2 d_{\mathrm{maj}} d_{\mathrm{min}}},
\]
where $F$ is the flux density of the component in jansky, $\nu$ is the observing frequency in gigahertz, and $d_{\mathrm{maj}}$ and $d_{\mathrm{min}}$ are the major and minor axes in milliarcseconds. This value corresponds to the observer’s frame. For the intrinsic rest-frame brightness temperature, we applied the correction
\[
T^\prime_B = T_B \frac{1+z}{\delta},
\]
with redshift $z$ and Doppler factor $\delta$. For typical Doppler factors of $\delta \sim 10$ (see Sect. \ref{sec:astrometry}), we obtain $T^\prime_B \sim 10^9$–$10^{10}\,\mathrm{K}$. These values are fully consistent with \citet{Kim2020} but remain significantly below both the equipartition and inverse Compton limits, suggesting an optically thin emission region or magnetically dominated plasma overall.

A recent statistical analysis of the 2017 EHT data by \citet{Roeder2025} revealed systematically low brightness temperatures across many sources, with 3C~279 being no exception. They find that intrinsic $T_B$ evolves with frequency as $\propto \nu^{-1}$ and increases with core separation as $\propto r^{0.95}$. Their study concludes that the classical Blandford–K\"{o}nigl jet model \citep{Blandford1979} is insufficient to explain the observations and that either bulk acceleration of the jet or a transfer of energy from magnetic fields to relativistic particles is required.

Several possible explanations can be considered. Ongoing internal acceleration within the elongated jet could, in principle, account for the observed values, which is also supported by MOJAVE monitoring \citep[e.g.,][]{Homan2015}. Another plausible interpretation for the low brightness temperatures of the jet components is that they reflect reduced opacity at 230\,GHz, where the observing frequency exceeds the turnover frequency. Indeed, ALMA observations of 3C~279 report a spectral curvature in addition to a steep spectrum toward higher frequencies \citep{Goddi2019, Goddi2021}, consistent with an optically thin regime. However, since ALMA does not resolve the core region, the measured spectrum is likely dominated by the optically thin jet emission ($\alpha \sim -0.5$), i.e., C2 is brighter than C0.

\subsection{Jet dynamics} \label{sec:astrometry}

\begin{figure*}
    \centering
    \begin{subfigure}[t]{0.32\textwidth}
        \centering
        \includegraphics[width=\linewidth]{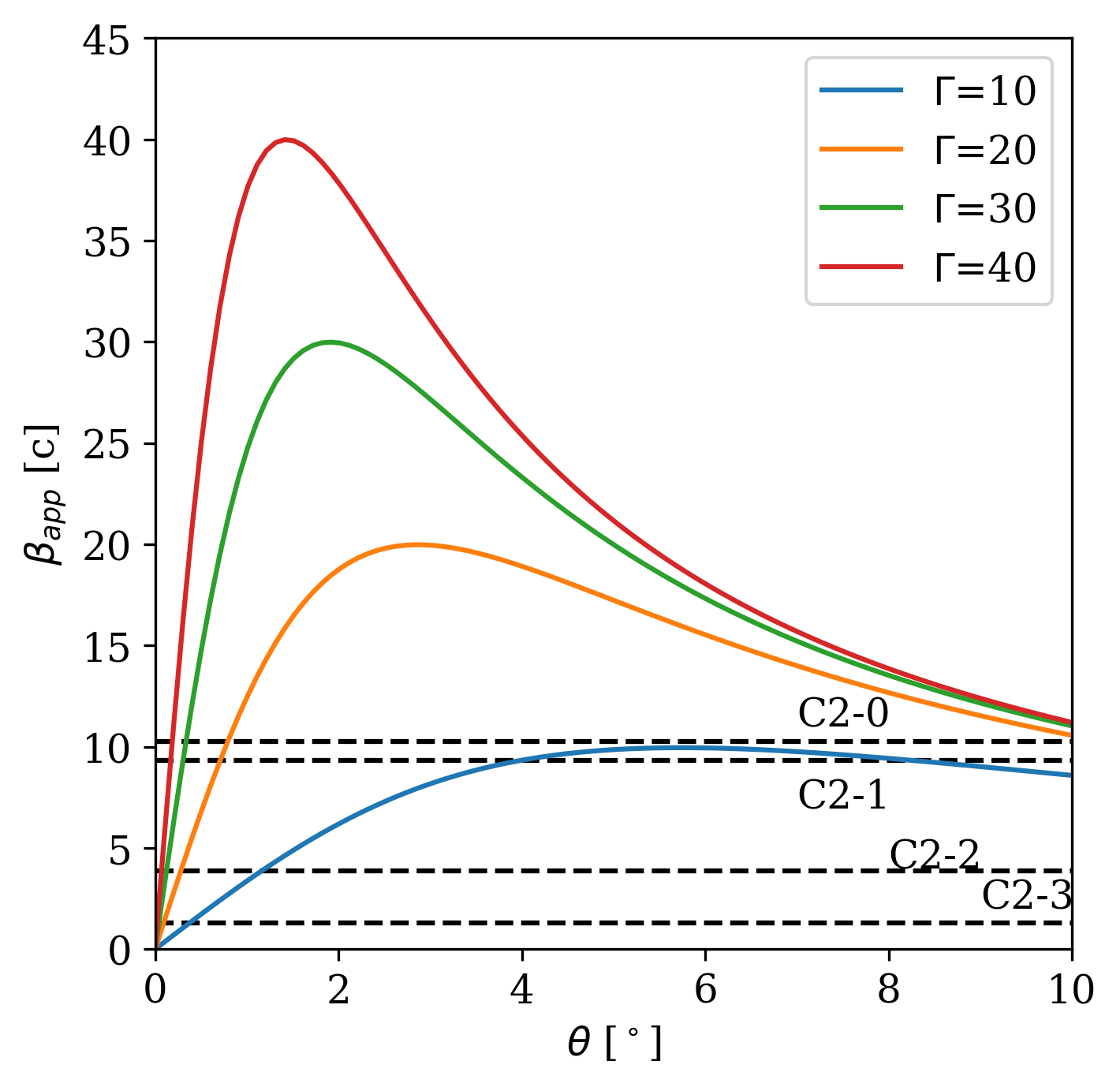}
        \caption{}
        \label{fig:theta_beta}
    \end{subfigure}
    \begin{subfigure}[t]{0.32\textwidth}
        \centering
        \includegraphics[width=\linewidth]{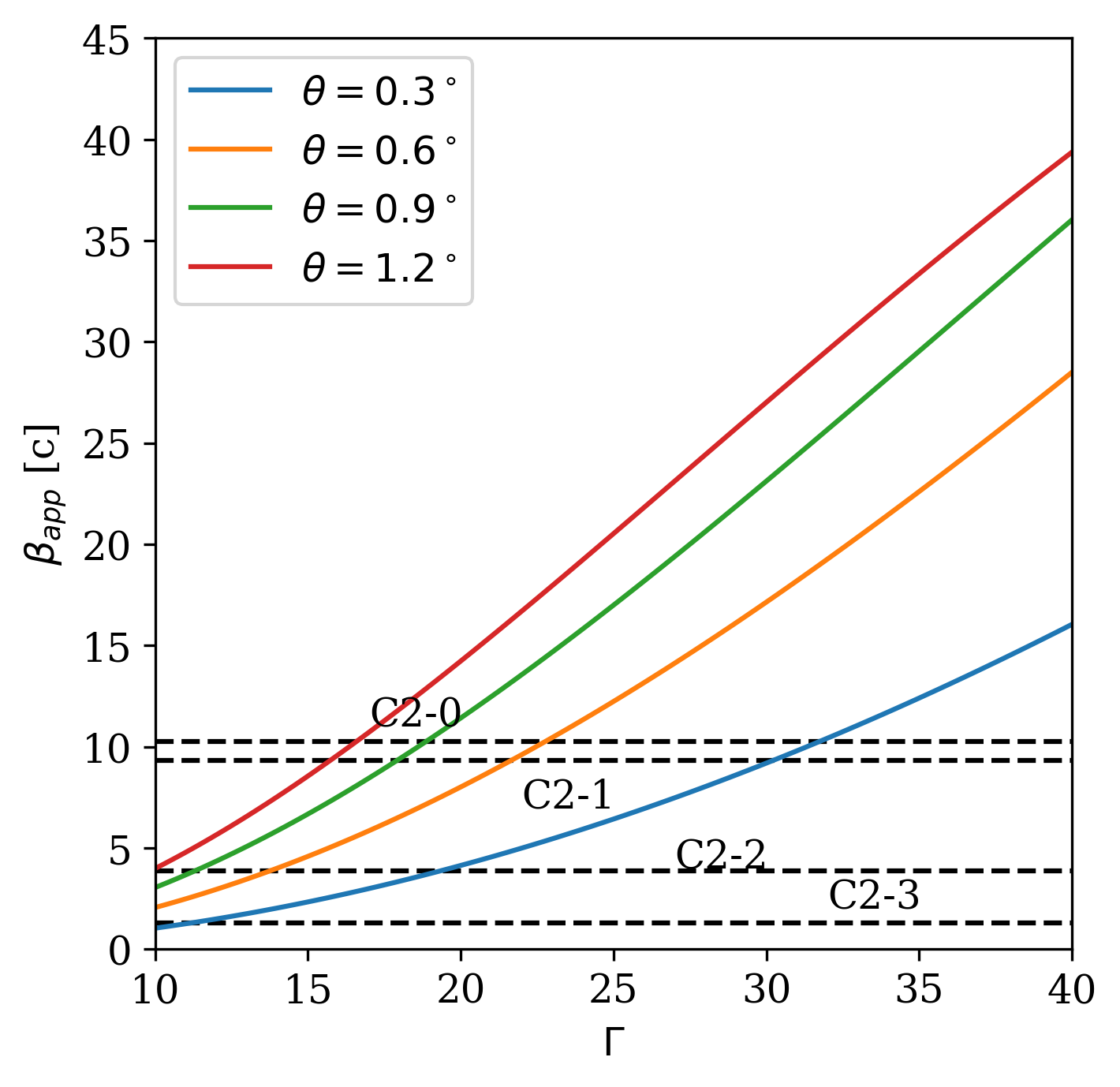}
        \caption{}
        \label{fig:gamma_beta}
    \end{subfigure}
    \begin{subfigure}[t]{0.32\textwidth}
        \centering
        \includegraphics[width=\linewidth]{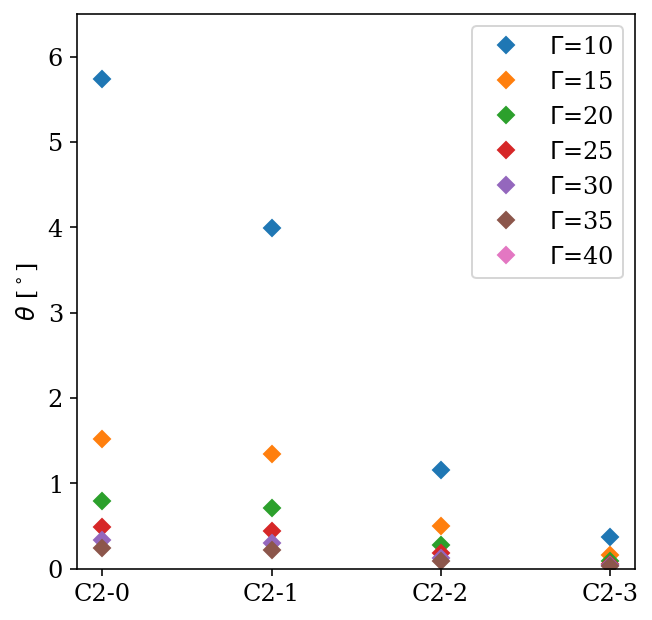}
        \caption{}
        \label{fig:viewing_angles}
    \end{subfigure}
    \caption{Comparison of apparent speed $\beta_{\rm{app}}$ and viewing angle constraints for 3C~279. Each panel shows a different parameterization: (a) Apparent speed as function of viewing angle for different bulk Lorentz factors compared to measured values as indicated by the dashed lines. (b) Apparent speed as function of the bulk Lorentz factor for different viewing angles compared to measured values. (c) Upper limits on viewing angles for components C2-0, C2-1, C2-2, and C2-3, depending on the assumed bulk Lorentz factor.}
    \label{fig:combined}
\end{figure*}

We measured C2's four subcomponents speeds using model fitting based on our band 3 results. However, we note that we obtain very similar modeling results independently from band 4 as well (see Appendix \ref{app:additional_reconstruction}). The motions of C2-1, C2-2, C2-3, and C3 align with the large-scale jet, while C2-0 exhibits a peculiar trajectory. The components' direction of motion does not align with the large-scale jet but rather with the bending of the jet. The measured apparent speeds are listed in Table~\ref{tab:speeds}.
The apparent speed $\beta_{\mathrm{app}}$ is related to the intrinsic speed $\beta$ through
\begin{align}
    \beta = \frac{\beta_{\mathrm{app}}}{\sin \theta + \beta_{\mathrm{app}} \cos \theta},
\end{align}
where $\theta$ denotes the viewing angle. The intrinsic speed is connected to the bulk Lorentz factor $\Gamma$ via
\begin{align}
    \Gamma = \frac{1}{\sqrt{1-\beta^2}},
\end{align}
and the Doppler factor $\delta$ is defined as
\begin{align}
    \delta = \frac{1}{\Gamma (1-\beta \cos \theta)}.
\end{align}

Figure~\ref{fig:theta_beta} illustrates the measured apparent speeds together with the dependence of $\beta_{\mathrm{app}}$ on the viewing angle for different values of $\Gamma$. The minimum Lorentz factor is limited by the observed apparent speed. For component C2-0, we derive a lower limit of $\Gamma = 10.3 \pm 0.5$. This estimate is consistent with the EHT measurements presented in \citet{Kim2020} and agrees with values reported by MOJAVE monitoring at 15\,GHz further downstream in the jet \citep{Homan2015}. The Doppler factors for $\Gamma = 10.3$ are $10 \pm 3$ for C2-0, $14 \pm 2$ for C2-1, $19 \pm 1$ for C2-2, and $20 \pm 1$ for C2-3.

\citet{Kim2020} argue that the acceleration zone of the jet in 3C~279 must be located near the kinematic origin, within the inner $\sim 20\,\mu\mathrm{as} \simeq 1600~\mathrm{r_g}$. Our measurements agree with this interpretation: the C2 components move at Lorentz factors comparable to the outer jet, consistent with acceleration occurring only close to C0. In this interpretation, the cluster of the C2 components would naturally represent master recollimation shock. This interpretation would also be consistent with C2's high brightness compared to the apparent jet apex, i.e., the kinematic origin C0. In fact, it is a common scenario in AGNs that the recollimation shock outshines the apparent jet apex \citep[e.g.][]{Agudo2012,Fromm2013,Hodgson2017}.

For C2-1, C2-2, and C2-3, we find apparent speeds that decrease with increasing separation from the kinematic origin. This can be explained by geometric projection effects, such as a bending of the jet toward the observer that reduces the viewing angle. By Occam's razor, this explanation is preferred, as it does not require invoking genuine jet deceleration and allows the components to share a similar bulk Lorentz factor, consistent with the interpretation proposed by \citet{Kim2020}. We note that \citet{Kim2020} report the non-radial motion for only two components. Here, we report the C2-0 component whose motion does not align with the jet's milliarcsecond-scale orientation.

Based on these observations, we can constrain the viewing angle for the different components. Assuming a minimum bulk Lorentz factor of $\Gamma \gtrsim 10.3$, we can determine the largest viewing angle that reproduces the apparent speeds of the components, in particular C2-3. Figure~\ref{fig:gamma_beta} shows the dependence of apparent speed on $\Gamma$ for different viewing angles $\theta$, while Fig.~\ref{fig:viewing_angles} presents the corresponding upper limits, listed in Table~\ref{tab:speeds}. The relative brightness ratios of the C2-x components can thus, at least partially, be explained by their differences in the viewing angle, resulting in different Doppler factors.

In this interpretation, the jet bends toward the observer, resulting in extremely small viewing angles, such as $\theta_{C2-3} < 0.3^\circ$, even smaller than previously reported for the large-scale jet \citep{Jorstad2017}. Consequently, 3C~279 (at least in April 2021) appears as one of the few blazars seen almost face-on. Such a small viewing angle would naturally produce a very wide opening angle close to C2-3. This is likely not recovered, as the sensitivity of EHT imaging to jet emission is limited. This interpretation is also supported by the EVPA distribution of 3C~279 at 43 GHz, for instance, in the BU-Blazar data of the source \citep[e.g.,][]{Jorstad2017}, which shows close to radially diverging geometry that is linked to small viewing angles in sources such as PKS~1510$-$089 \citep{Homan2002}. However, other plausible phenomena could similarly produce overly bright regions, such as a major recollimation shock occurring at the end of the acceleration zone (which, in this interpretation, extends from C0 to C2-x). However, without additional bending geometry, this interpretation does not explain the observed deceleration in apparent speed from C2-0 toward C2-3 on its own. In fact, the most plausible scenario might be a combination of a bent jet geometry and shocks.

Furthermore, we note that the C0 component appears as an elongated parabola elongated in the east-west direction and opening toward C2 (cf., Fig. \ref{fig:e21e13_b3} and the left panel in Fig. \ref{fig:components}). Assuming C0 represents the jet apex would produce an unusually high apparent opening angle of $130^\circ \pm 10^\circ$, which quickly collimates to a cylindrical shape at C2. For the small viewing angles estimated in this analysis, i.e., for $\theta < 5.3^\circ$ (as estimated for C2-0), this corresponds to an intrinsic opening angle $<13^\circ$.

\begin{table*}
\centering
\caption{Kinematic and brightness temperature constraints for components C2-0 to C2-3 measured between April 13 and April 18.}
\label{tab:speeds}
\begin{tabular}{lcccc}
\toprule
Component & $\beta_{app}$ & Projected speed [$\mu \mathrm{as}/\mathrm{day}$]& $\theta$ [$^\circ$] & Rest Frame Brightness Temperature [$10^9\,\mathrm{K}$] \\
\midrule
C2-0 & $10.2 \pm 4.3$ & $1.2 \pm 0.5$ & $<5.3$ & $<2.1$ \\
C2-1 & $9.3 \pm 1.6$ & $1.1 \pm 0.1$ & $<3.6$ & $<11.4$ \\
C2-2 & $3.9 \pm 2.3$ & $0.4 \pm 0.1$ & $<1.1$ & $<17.4$ \\
C2-3 & $1.3 \pm 1.1$ & $0.2 \pm 0.1$ & $<0.3$ & $<4.1$ \\
\bottomrule
\end{tabular}
\end{table*}

\subsection{Multiyear evolution of the innermost region of the 3C~279 jet} \label{sec:sourceEvolution}
\begin{figure*}
    \centering
    \includegraphics[width=\textwidth]{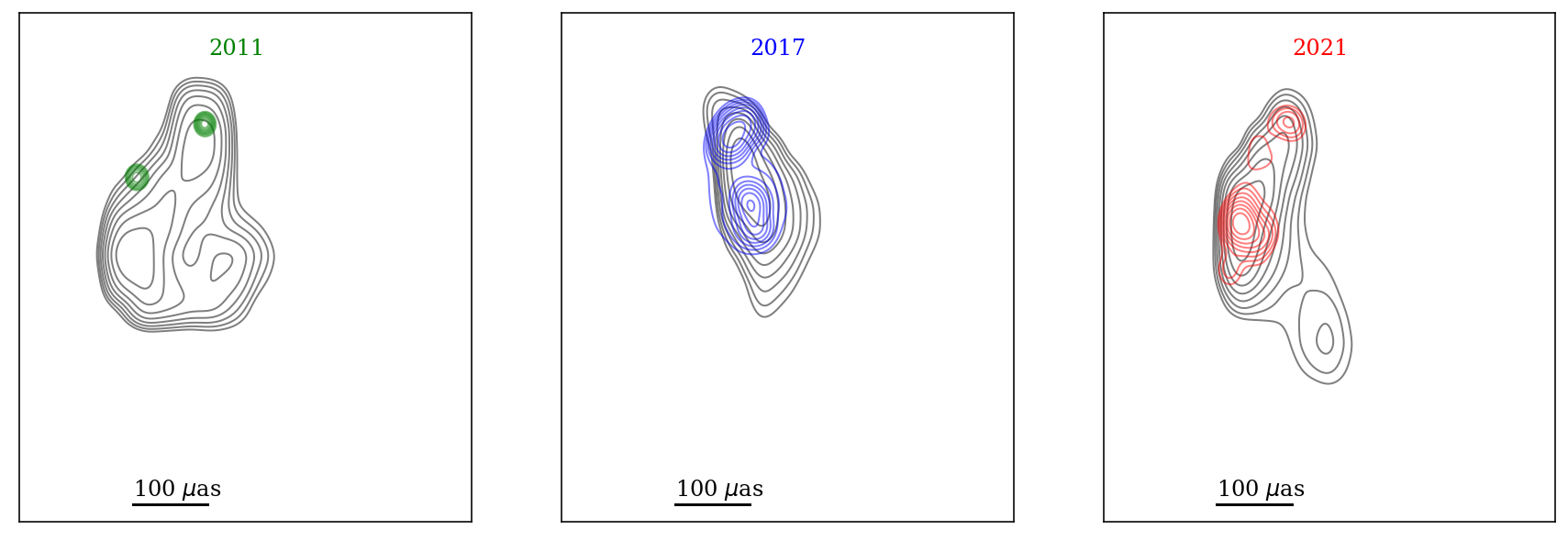}
    \caption{43\,GHz images of 3C~279 (black), obtained by the University of Boston blazar monitoring program at epochs closest in time to the three EHT at 230\,GHz (2011: green, 2017: blue, and 2021: red) with the EHT images overplotted. The EHT data were taken on April 11, 2017 and April 13, 2021. The 2011 EHT results represent a combined fit of data obtained between March 29 and April 4, 2011. The corresponding 43 GHz epochs are from April 21, 2011, April 16, 2017, and April 5, 2021. The reconstructions were performed with the super-resolving dynamic imaging technique \texttt{kine}. The contour levels increase by a factor of 2 from $0.19\%$ to $50\%$ of the respective peak flux. The peak fluxes at 43 GHz are 1.5,0.7,1.9 mJy/$\mu \mathrm{as}^2$, and 6.4,1.7 and 3.5 mJy/$\mu \mathrm{as}^2$ at 230 GHz.}
    \label{fig:multiyear_43}
\end{figure*}

\autoref{fig:multiyear} presents a comparison of all EHT and pre‑EHT observations of the source. 
The 2017 EHT image of 3C~279 revealed an elongated core structure oriented roughly perpendicular to the large‑scale jet. While the 2017 and 2021 images show roughly similar morphologies, a pronounced structural change is evident when compared to the 2011 and 2021 data. These differences cannot be attributed solely to the day‑to‑day variability observed within the individual 2017 and 2021 epochs, as the relative position angles of the image components also evolve.
If the kinematic origin location (component C0) is assumed to remain fixed, the observed changes imply a reconfiguration of the brightest emission features at the jet base. This may reflect an intrinsic alteration of the jet morphology or plasma content, a change in viewing geometry leading to modified Doppler boosting, or a combination of these effects.
In Fig. \ref{fig:multiyear_43}, we show 43 GHz monitoring results on 3C~279 from epochs closest in time to the EHT and proto-EHT epochs of 2011, 2017 and 2021. These images were created by reanalyzing the datasets with the dynamic imaging code \texttt{kine}, which achieves significant super-resolution. We note that the position angle swing observed at 230\,GHz is mirrored at smaller frequencies as well.
Intriguingly, the evolution of the position angle appears confined to the innermost regions (on approximate parsec scales when de‑projected) and contrasts with the apparently straight jet observed on larger scales. This suggests that the jet roughly 100\,$\mu\mathrm{as}$ downstream from the kinematic origin (close to the structural bent) undergoes structural evolution on comparatively short, year‑level timescales, whereas the outer jet follows the more stable, precleared channel established over much longer periods.

\section{Discussion and conclusions}
The 2021 EHT observations of 3C~279 reveal a series of compact features elongated in a north–south direction, nearly orthogonal to the large-scale jet axis traced at longer wavelengths. This morphology, consistent with earlier EHT results from 2017 and 2011 1.3\,mm VLBI imaging \citep{Lu2013}, along with 86 GHz imaging and 43 GHz monitoring of 3C~279, suggests that the observed misalignment is not a transient imaging artifact but rather an intrinsic property of the innermost jet. The recurrence of this feature across multiple epochs combined with the apparent evolution of the overall jet-morphology points to a persistent structural characteristic of the source on microarcsecond scales.
We constrain the bulk Lorentz factor at values $\Gamma \gtrsim 10$ within the inner $\sim 5000$ gravitational radii (de-projected) from the kinematic origin, indicating that acceleration occurs near the kinematic origin. The first downstream jet component is significantly brighter than the kinematic origin, potentially representing the master recollimation shock. Moreover, we find very small viewing angles for 3C~279 and a bent structure.

Such bent or transversely extended inner jet structures are not unique to 3C~279. Similar phenomena have been reported in other prominent blazars, including OJ~287 \citep{Agudo2012,Hodgson2017}, 3C~345 \citep{Lobanov2005,Traianou2024}, NRAO530 \citep{Lisakov2025}, 3C84 \citep{Paraschos2024}, PKS 2136+141 \citep{Savolainen2006}, and CTA~102 \citep{Fromm2013}. The convergence of evidence across sources suggests that relativistic jets may commonly undergo geometric distortions on (sub-)parsec scales, in some cases linked to episodes of enhanced activity \citep[e.g.,][]{Traianou2024}. \citet{Principe2026} report a record UV–optical flare with strong polarization variability occurring a couple of weeks prior to the 2017 EHT observations, followed by a gamma-ray flaring activity that declined before the end of the 2017 EHT observing period, potentially linked to the strongly bent jet appearance of 3C~279 in that year.

This raises the possibility of a shared physical mechanism underlying these behaviors.
Several scenarios have been proposed to explain the jet bending at these scales. In the context of thin-disk accretion, i.e., applicable to 3C~279, jet precession may arise from Lense-Thirring torques in a misaligned accretion flow to drive the jet precession \citep{Liska2018_precession, Liska2021_diskTearingBardeen}. 
However, jet precession, or more generically dynamic jet evolution, is not exclusive to quasars and is also reported in other accretion regimes. Specifically, it has also been reported for thick, radiatively inefficient accretion flows (RIAFs), such as M87* \citep[e.g.,][]{Walker2018, Cui2023}. Recent general relativistic magnetohydrodynamic (GRMHD) simulations suggest that Lense-Thirring-driven jet precession is unlikely, or at best transient, in RIAFs \citep[]{Ressler2023,Chatterjee2025}. Since it also requires an initial BH-disk misalignment, this mechanism alone cannot account for all observed cases of jet precession or jet evolution, implying that additional processes may be involved.
One additional possibility is jet evolution arising from magnetic-eruption events triggered by reconnection near the event horizon, which can reconfigure both the inner accretion flow and the jet base \citep[e.g.,][]{Ripperda2022,Davelaar2023,Galishnikova2025,Ressler2025,Lalakos2025}, and may also occur in truncated thin disks \citep{Liska2022}. Another is kink instability arising from the interaction between the jet and the surrounding ambient material \citep[e.g.,][]{Tchekhovskoy2016kinky,Bromberg2016,Sironi2021,Ressler2021_kinkUnstable,Lalakos2024,Lalakos2025}. A further scenario involves the growth of standing Kelvin–Helmholtz (K-H) instabilities within the jet \citep[e.g.,][]{Chow2023}, a process proposed to explain the evolution and helical structure of the 3C~279 jet \citep[e.g.,][]{Fuentes2023}, and in other jet sources such as M87* \citep[e.g.][]{Walker2018, Nikonov2023}. The latter two processes (kink and K--H instability) have been demonstrated in simulations of RIAF accretion flows, but their presence -- or absence -- in the (truncated) thin-disk regime applicable in the 3C~279 case remains less certain. Nevertheless, their accretion independence suggests that they may operate there as well.

\cite{Qian2019} suggest the presence of a binary supermassive black hole system in 3C~279, which could imprint quasiperiodic changes in jet orientation with a $\sim 22~\mathrm{yr}$ period. Such models have been most prominently discussed in the context of OJ287 \citep[e.g.,][]{Agudo2012,Britzen2018} but also in other sources such as M81*, 3C345, PKS 2131$-$021, and PKS J0805$-$0111 \citep[e.g.,][]{Lobanov2005, vonFellenberg2023M81,Jiang2023,Sullivan2026}. In such an interpretation, the binary may induce jet precession directly through the reflex motion of primary \citep[e.g.,][]{Begelman1980}, spin-orbit coupling \citep[e.g.,][]{Ressler2024}, or perturbations of the accretion flow through the secondary \citep[e.g.,][]{Ressler2025}. 3C~279 shows significant position angle variability in the innermost jet across multiple years. However, the limited number of high-resolution EHT epochs (2011, 2017, and 2021) still prevents the robust testing of such models. Thus, the potential influence of binary dynamics or secular precession cannot yet be ruled out. Nevertheless, it appears to be the unlikely cause of precession in \textit{all systems}, and in the case of 3C~279, no specific evidence points to the presence of a binary companion. 

Despite the concrete evidence for (sub)-pc-scale evolution of jets on short timescales ($\mathcal{O}(10\,\mathrm{yr})$), the systems under discussion span a wide range of physical scales, with black hole masses of $10^8$--$10^{10}\,\mathrm{M_\odot}$, magnetic field strengths of $B \approx 10^0$--$10^4\,\mathrm{G}$, and accretion rates of $\dot{m} \approx 10^{-5}$--$10^{-2}\,\dot{m}_{\mathrm{edd}}$. While the absence of longer periods is a trivial selection bias due to the limited observation time span, this diversity in scales and accretion regimes poses a significant challenge for explaining the relatively short jet-modulation timescales observed. Still, given that large-scale magnetohydrodynamic simulations of accreting black hole systems routinely produce rich, time-dependent jet-ambient medium interactions and that analogous behavior is observed across a wide range of astrophysical jets, it seems plausible that at least part of the observed variability arises from the jet propagating through an inhomogeneous, evolving environment. 
Given the excellent imaging and relative astrometric performance of the EHT, continued monitoring of the source, together with complementary VLBI arrays, will be essential for disentangling these possibilities and constraining the physical mechanisms responsible for jet variability.

\bibliographystyle{aa}
\bibliography{references}{}

\appendix

\section{Acknowledgments}
Figure 1 is only available in electronic form at the CDS via anonymous ftp to cdsarc.u-strasbg.fr (130.79.128.5) or via \url{http://cdsweb.u-strasbg.fr/cgi-bin/qcat?J/A+A/}.
\begin{acknowledgements}
    This research has made use of data obtained with the Global Millimeter VLBI Array (GMVA), which consists of telescopes operated by the MPIfR, IRAM, Onsala, Metsahovi, Yebes, the Korean VLBI Network, the Greenland Telescope, the Green Bank Observatory and the Very Long Baseline Array (VLBA). The VLBA and the GBT are facilities of the National Science Foundation operated under cooperative agreement by Associated Universities, Inc. The data were correlated at the correlator of the MPIfR in Bonn, Germany.
    This study makes use of VLBA data from the VLBA-BU Blazar Monitoring Program (BEAM-ME and VLBA-BU-BLAZAR; \url{http://www.bu.edu/blazars/BEAM-ME.html}), funded by NASA through the Fermi Guest Investigator Program. The VLBA is an instrument of the National Radio Astronomy Observatory. The National Radio Astronomy Observatory is a facility of the National Science Foundation operated by Associated Universities, Inc.
    HM acknowledges the support of National Radio Astronomy Observatory through the Jansky Fellowship.
    SDvF gratefully acknowledges the support of the Alexander von Humboldt Foundation through a Feodor Lynen Fellowship and thanks CITA for their hospitality and collaboration.
The Event Horizon Telescope Collaboration thanks the following
organizations and programs: the Academia Sinica; the Research Council of Finland (project 362572); the Agencia Nacional de Investigaci\'{o}n y Desarrollo (ANID), Chile via NCN$19\_058$ (TITANs), Fondecyt 1221421 and 11251078, and BASAL FB210003; the Alexander
von Humboldt Stiftung (including the Feodor Lynen Fellowship); an Alfred P. Sloan Research Fellowship;
the ALMA North America Development Fund; the Astrophysics and High Energy Physics programme by MCIN (with funding from European Union NextGenerationEU, PRTR-C17I1); the Black Hole Initiative, which is funded by grants from the John Templeton Foundation (60477, 61497, 62286) and the Gordon and Betty Moore Foundation (Grants GBMF-8273, GBMF12987) -- although the opinions expressed in this work are those of the authors and do not necessarily reflect the views of these Foundations; 
the Brinson Foundation; the Canada Research Chairs (CRC) program; 
the University of Toronto Eric and Wendy Schmidt AI in Science Postdoctoral Fellowship Program, a program of Schmidt Sciences;
the Natural Sciences \& Engineering Research Council of Canada (NSERC);
the Ontario Research Fund - Research Excellence (Project Number RE012-045);
Chandra DD7-18089X and TM6-17006X; the China Scholarship
Council; the China Postdoctoral Science Foundation fellowships (2020M671266, 2022M712084); ANID through Fondecyt Postdoctorado (project 3250762); Conicyt through Fondecyt Postdoctorado (project 3220195); Consejo Nacional de Humanidades, Ciencia y Tecnología (CONAHCYT, Mexico, projects U0004-246083, U0004-259839, F0003-272050, M0037-279006, F0003-281692, 104497, 275201, 263356, CBF2023-2024-1102, 257435, Ph.D. Scholarship 963437); the Colfuturo Scholarship; the Consejo Superior de Investigaciones 
Cient\'{i}ficas (grant 2019AEP112);
the Delaney Family via the Delaney Family John A.
Wheeler Chair at Perimeter Institute; Dirección General de Asuntos del Personal Académico-Universidad Nacional Autónoma de México (DGAPA-UNAM, projects IN112820 and IN108324); the Dutch Research Council (NWO) for the VICI award (grant 639.043.513), the grant OCENW.KLEIN.113, and the Dutch Black Hole Consortium (with project No. NWA 1292.19.202) of the research programme the National Science Agenda; the Dutch National Supercomputers, Cartesius and Snellius  (NWO grant 2021.013); 
the EACOA Fellowship awarded by the East Asia Core
Observatories Association, which consists of the Academia Sinica Institute of Astronomy and Astrophysics, the National Astronomical Observatory of Japan, Center for Astronomical Mega-Science,
Chinese Academy of Sciences, and the Korea Astronomy and Space Science Institute; 
the European Research Council (ERC) Synergy Grant ``BlackHoleCam: Imaging the Event Horizon of Black Holes'' (grant 610058), Synergy Grant ``BlackHolistic:  Colour Movies of Black Holes:
Understanding Black Hole Astrophysics from the Event Horizon to Galactic Scales'' (grant 10107164), and Horizon ERC Grants 2021 program under grant agreement No. 101040021; 
the European Union Horizon 2020
research and innovation programme under grant agreements
BlackHolistic (No. 101071643), 
RadioNet (No. 730562), 
M2FINDERS (No. 101018682); the European Research Council for advanced grant ``JETSET: Launching, propagation and 
emission of relativistic jets from binary mergers and across mass scales'' (grant No. 884631); the European Horizon Europe staff exchange (SE) programme HORIZON-MSCA-2021-SE-01 grant NewFunFiCO (No. 10108625); the Horizon ERC Grants 2021 programme under grant agreement No. 101040021; the FAPESP (Funda\c{c}\~ao de Amparo \'a Pesquisa do Estado de S\~ao Paulo) under grant 2021/01183-8; the Fondes de Recherche Nature et Technologies (FRQNT); the Fondo CAS-ANID folio CAS220010; the Generalitat Valenciana (grants APOSTD/2018/177 and  ASFAE/2022/018) and
GenT Program (project CIDEGENT/2018/021);
the Hellenic Foundation for Research and Innovation (ELIDEK) under Grant No. 23698; 
the Gordon and Betty Moore Foundation (GBMF-3561, GBMF-5278, GBMF-10423);   
the Institute for Advanced Study; the ICSC – Centro Nazionale di Ricerca in High Performance Computing, Big Data and Quantum Computing, funded by European Union – NextGenerationEU; the Istituto Nazionale di Fisica
Nucleare (INFN) sezione di Napoli, iniziative specifiche
TEONGRAV; the Research Foundation -- Flanders (FWO) Postdoctoral Fellowship 1255226N; 
the International Max Planck Research
School for Astronomy and Astrophysics at the
Universities of Bonn and Cologne; the Italian Ministry of University and Research (MUR)– Project CUP F53D23001260001, funded by the European Union – NextGenerationEU; 
Deutsche Forschungsgemeinschaft (DFG, German Research Foundation) as part of the DFG Research Unit FOR5195 -– project number 443220636;
Joint Columbia/Flatiron Postdoctoral Fellowship (research at the Flatiron Institute is supported by the Simons Foundation); 
the Japan Ministry of Education, Culture, Sports, Science and Technology (MEXT; grant JPMXP1020200109); 
the Japan Society for the Promotion of Science (JSPS) Grant-in-Aid for JSPS
Research Fellowship (JP17J08829); the Joint Institute for Computational Fundamental Science, Japan; the Key Research
Program of Frontier Sciences, Chinese Academy of
Sciences (CAS, grants QYZDJ-SSW-SLH057, QYZDJSSW-SYS008, ZDBS-LY-SLH011); 
the Leverhulme Trust Early Career Research
Fellowship; the Max-Planck-Gesellschaft (MPG);
the Max Planck Partner Group of the MPG and the
CAS; the MEXT/JSPS KAKENHI (grants 18KK0090, JP21H01137,
JP18H03721, JP18K13594, 18K03709, JP19K14761, 18H01245, 25120007, 19H01943, 21H01137, 21H04488, 22H00157, 23K03453, 24KJ0773); the MICINN Research Projects PID2019-108995GB-C22, PID2022-140888NB-C22; the MIT International Science
and Technology Initiatives (MISTI) Funds; 
the Ministry of Science and Technology (MOST) of Taiwan (103-2119-M-001-010-MY2, 105-2112-M-001-025-MY3, 105-2119-M-001-042, 106-2112-M-001-011, 106-2119-M-001-013, 106-2119-M-001-027, 106-2923-M-001-005, 107-2119-M-001-017, 107-2119-M-001-020, 107-2119-M-001-041, 107-2119-M-110-005, 107-2923-M-001-009, 108-2112-M-001-048, 108-2112-M-001-051, 108-2923-M-001-002, 109-2112-M-001-025, 109-2124-M-001-005, 109-2923-M-001-001, 
110-2112-M-001-033, 110-2124-M-001-007 and 110-2923-M-001-001); the National Science and Technology Council (NSTC) of Taiwan
(111-2124-M-001-005, 112-2124-M-001-014,  112-2112-M-003-010-MY3, and 113-2124-M-001-008);
the Ministry of Education (MoE) of Taiwan Yushan Young Scholar Program;
the Physics Division, National Center for Theoretical Sciences of Taiwan;
the National Aeronautics and
Space Administration (NASA, Fermi Guest Investigator
grant 
80NSSC23K1508, NASA Astrophysics Theory Program grant 80NSSC20K0527, NASA NuSTAR award 
80NSSC20K0645); NASA Hubble Fellowship Program Einstein Fellowship;
NASA Hubble Fellowship 
grants HST-HF2-51431.001-A, HST-HF2-51482.001-A, HST-HF2-51539.001-A, HST-HF2-51552.001A awarded 
by the Space Telescope Science Institute, which is operated by the Association of Universities for 
Research in Astronomy, Inc., for NASA, under contract NAS5-26555; 
the National Institute of Natural Sciences (NINS) of Japan; the National
Key Research and Development Program of China
(grant 2016YFA0400704, 2017YFA0402703, 2016YFA0400702); the National Science and Technology Council (NSTC, grants NSTC 111-2112-M-001 -041, NSTC 111-2124-M-001-005, NSTC 112-2124-M-001-014); the US National Science Foundation (NSF, grants AST-0096454,
AST-0352953, AST-0521233, AST-0705062, AST-0905844, AST-0922984, AST-1126433, OIA-1126433, AST-1140030,
DGE-1144085, AST-1207704, AST-1207730, AST-1207752, MRI-1228509, OPP-1248097, AST-1310896, AST-1440254, 
AST-1555365, AST-1614868, AST-1615796, AST-1715061, AST-1716327,  AST-1726637, 
OISE-1743747, AST-1743747, AST-1816420, AST-1935980, AST-1952099, AST-2034306,  AST-2205908, AST-2307887, AST-2407810, AST-2535855); 
NSF Astronomy and Astrophysics Postdoctoral Fellowship (AST-1903847); 
the NSF Graduate Research Fellowship (DGE 2140743);
the Natural Science Foundation of China (grants 11650110427, 10625314, 11721303, 11725312, 11873028, 11933007, 11991052, 11991053, 12192220, 12192223, 12273022, 12325302, 12303021); 
the AWS Impact Computing Project at the Harvard Data Science Initiative (award no. A61166); 
the Natural Sciences and Engineering Research Council of
Canada (NSERC);
the Korea Aerospace Administration (KASA) (grant no. RS-2026-25587698); 
the National Research Foundation of Korea (the Global PhD Fellowship Grant: grants NRF-2015H1A2A1033752; the Korea Research Fellowship Program: NRF-2015H1D3A1066561; Brain Pool Program: RS-2024-00407499;  Basic Research Support Grant 2019R1F1A1059721, 2021R1A6A3A01086420, 2022R1C1C1005255, RS-2022-NR071771, RS-2025-16067786, RS-2025-02214038); the POSCO Science Fellowship of the POSCO TJ Park Foundation; the Global University 30 Project Fund of Kyungpook National University in 2026; the Global-Learning \& Academic research institution for Master's \textperiodcentered~PhD students, and Postdocs(G-LAMP) Program of the National Research Foundation of Korea (NRF) grant funded by the Ministry of Education (grants RS-2023-00301914, RS-2025-25442355); NOIRLab, which is managed by the Association of Universities for Research in Astronomy (AURA) under a cooperative agreement with the National Science Foundation; 
the A.G. Leventis Foundation; 
Onsala Space Observatory (OSO) national infrastructure, for the provisioning
of its facilities/observational support (OSO receives funding through the Swedish Research Council under grant 2017-00648);  the Perimeter Institute for Theoretical Physics (research at Perimeter Institute is supported by the Government of Canada through the Department of Innovation, Science and Economic Development and by the Province of Ontario through the Ministry of Research, Innovation and Science); the Portuguese Foundation for Science and Technology (FCT) grants (Individual CEEC program – 5th edition, CIDMA
through the FCT Multi-Annual Financing Program for R\&D Units UID/04106, CERN/FIS-PAR/0024/2021, 2022.04560.PTDC); the Princeton Gravity Initiative; the Spanish Ministerio de Ciencia, Innovaci\'{o}n  y Universidades (grants PID2022-140888NB-C21, PID2022-140888NB-C22, PID2023-147883NB-C21, RYC2023-042988-I); the Severo Ochoa grant CEX2021-001131-S funded by MICIU/AEI/10.13039/501100011033; The European Union’s Horizon Europe research and innovation program under grant agreement No. 101093934 (RADIOBLOCKS); The European Union “NextGenerationEU”, the Recovery, Transformation and Resilience Plan, the CUII of the Andalusian Regional Government and the Spanish CSIC through grant AST22\_00001\_Subproject\_10; ``la Caixa'' Foundation (ID 100010434) through fellowship codes LCF/BQ/DI22/11940027 and LCF/BQ/DI22/11940030 and ; 
the University of Pretoria for financial aid in the provision of the new 
Cluster Server nodes and SuperMicro (USA) for a SEEDING GRANT approved toward these 
nodes in 2020; the Shanghai Municipality orientation program of basic research for international scientists (grant no. 22JC1410600); 
the Shanghai Pilot Program for Basic Research, Chinese Academy of Science, 
Shanghai Branch (JCYJ-SHFY-2021-013); the Simons Foundation (grant 00001470); the Spanish Ministry for Science and Innovation grant CEX2021-001131-S funded by MCIN/AEI/10.13039/501100011033; the Spinoza Prize SPI 78-409; the South African Research Chairs Initiative, through the 
South African Radio Astronomy Observatory (SARAO, grant ID 77948),  which is a facility of the National 
Research Foundation (NRF), an agency of the Department of Science and Innovation (DSI) of South Africa; the Swedish Research Council (VR); the Taplin Fellowship; the Toray Science Foundation; the UK Science and Technology Facilities Council (grant no. ST/X508329/1); the US Department of Energy (USDOE) through the Los Alamos National
Laboratory (operated by Triad National Security,
LLC, for the National Nuclear Security Administration
of the USDOE, contract 89233218CNA000001); and the YCAA Prize Postdoctoral Fellowship. This work was also supported by the National Research Foundation of Korea (NRF) grant funded by the Korea government(MSIT) (RS-2024-00449206). We acknowledge support from the Coordenação de Aperfeiçoamento de Pessoal de Nível Superior (CAPES) of Brazil through PROEX grant number 88887.845378/2023-00. We acknowledge financial support from Millenium Nucleus NCN23\_002 (TITANs) and Comité Mixto ESO-Chile.

We thank
the staff at the participating observatories, correlation
centers, and institutions for their enthusiastic support.
This paper makes use of the following ALMA data:
ADS/JAO.ALMA\#2017.1.00841.V and ADS/JAO.ALMA\#2019.1.01797.V.
ALMA is a partnership
of the European Southern Observatory (ESO;
Europe, representing its member states), NSF, and
National Institutes of Natural Sciences of Japan, together
with National Research Council (Canada), Ministry
of Science and Technology (MOST; Taiwan),
Academia Sinica Institute of Astronomy and Astrophysics
(ASIAA; Taiwan), and Korea Astronomy and
Space Science Institute (KASI; Republic of Korea), in
cooperation with the Republic of Chile. The Joint
ALMA Observatory is operated by ESO, Associated
Universities, Inc. (AUI)/NRAO, and the National Astronomical
Observatory of Japan (NAOJ).
The National Radio Astronomy Observatory (NRAO) and Green Bank Observatory are facilities of the U.S. National Science Foundation (NSF) operated under cooperative agreement by AUI.
This research used resources of the Oak Ridge Leadership Computing Facility at the Oak Ridge National
Laboratory, which is supported by the Office of Science of the U.S. Department of Energy under contract
No. DE-AC05-00OR22725; the ASTROVIVES FEDER infrastructure, with project code IDIFEDER-2021-086; the computing cluster of Shanghai VLBI correlator supported by the Special Fund 
for Astronomy from the Ministry of Finance in China;  
We also thank the Center for Computational Astrophysics, National Astronomical Observatory of Japan. This work was supported by FAPESP (Fundacao de Amparo a Pesquisa do Estado de Sao Paulo) under grant 2021/01183-8.

APEX is a collaboration between the
Max-Planck-Institut f{\"u}r Radioastronomie (Germany),
ESO, and the Onsala Space Observatory (Sweden). The
SMA is a joint project between the SAO and ASIAA
and is funded by the Smithsonian Institution and the
Academia Sinica. The JCMT is operated by the East
Asian Observatory on behalf of the NAOJ, ASIAA, and
KASI, as well as the Ministry of Finance of China, Chinese
Academy of Sciences, and the National Key Research and Development
Program (No. 2017YFA0402700) of China
and Natural Science Foundation of China grant 11873028.
Additional funding support for the JCMT is provided by the Science
and Technologies Facility Council (UK) and participating
universities in the UK and Canada. 
The LMT is a project operated by the Instituto Nacional
de Astr\'{o}fisica, \'{O}ptica, y Electr\'{o}nica (Mexico) and the
University of Massachusetts at Amherst (USA).
The IRAM 30 m telescope on Pico Veleta, Spain and the NOEMA interferometer on Plateau de Bure,
France are operated by IRAM and supported by CNRS (Centre National de la Recherche Scientifique, France), MPG (Max-Planck-Gesellschaft, Germany), and IGN (Instituto Geográfico Nacional, Spain).
The SMT is operated by the Arizona
Radio Observatory, a part of the Steward Observatory
of the University of Arizona, with financial support of
operations from the State of Arizona and financial support
for instrumentation development from the NSF.
Support for SPT participation in the EHT is provided by the National Science Foundation through award OPP-1852617 
to the University of Chicago. Partial support is also 
provided by the Kavli Institute of Cosmological Physics at the University of Chicago. The SPT hydrogen maser was 
provided on loan from the GLT, courtesy of ASIAA.

This work used the
Extreme Science and Engineering Discovery Environment
(XSEDE), supported by NSF grant ACI-1548562,
and CyVerse, supported by NSF grants DBI-0735191,
DBI-1265383, and DBI-1743442. XSEDE Stampede2 resource
at TACC was allocated through TG-AST170024
and TG-AST080026N. XSEDE JetStream resource at
PTI and TACC was allocated through AST170028.
This research is part of the Frontera computing project at the Texas Advanced 
Computing Center through the Frontera Large-Scale Community Partnerships allocation
AST20023. Frontera is made possible by National Science Foundation award OAC-1818253.
This research was done using services provided by the OSG Consortium~\citep{osg07,osg09}, which is supported by the National Science Foundation award Nos. 2030508 and 1836650.
Additional work used ABACUS2.0, which is part of the eScience center at Southern Denmark University, and the Kultrun Astronomy Hybrid Cluster (projects Conicyt Programa de Astronomia Fondo Quimal QUIMAL170001, Conicyt PIA ACT172033, Fondecyt Iniciacion 11170268, Quimal 220002). 
Simulations were also performed on the SuperMUC cluster at the LRZ in Garching, 
on the LOEWE cluster in CSC in Frankfurt, on the HazelHen cluster at the HLRS in Stuttgart, 
and on the Pi2.0 and Siyuan Mark-I at Shanghai Jiao Tong University.
The computer resources of the Finnish IT Center for Science (CSC) and the Finnish Computing 
Competence Infrastructure (FCCI) project are acknowledged. This
research was enabled in part by support provided
by Compute Ontario (http://computeontario.ca), Calcul
Quebec (http://www.calculquebec.ca), and the Digital Research Alliance of Canada (https://alliancecan.ca/en).

The EHTC has
received generous donations of FPGA chips from Xilinx
Inc., under the Xilinx University Program. The EHTC
has benefited from technology shared under open-source
license by the Collaboration for Astronomy Signal Processing
and Electronics Research (CASPER). The EHT
project is grateful to T4Science and Microsemi for their
assistance with hydrogen masers. This research has
made use of NASA's Astrophysics Data System. We
gratefully acknowledge the support provided by the extended
staff of the ALMA, from the inception of
the ALMA Phasing Project through the observational
campaigns of 2017 and 2018. We would like to thank
A. Deller and W. Brisken for EHT-specific support with
the use of DiFX. We thank Martin Shepherd for the addition of extra features in the Difmap software 
that were used for the CLEAN imaging results presented in this paper.
We acknowledge the significance that
Maunakea, where the SMA and JCMT EHT stations
are located, has for the indigenous Hawaiian people.
\end{acknowledgements}

\section{Additional reconstructions, model fits, and u-v coverage}\label{app:additional_reconstruction}

\begin{figure}
    \centering
    \includegraphics[width=\linewidth]{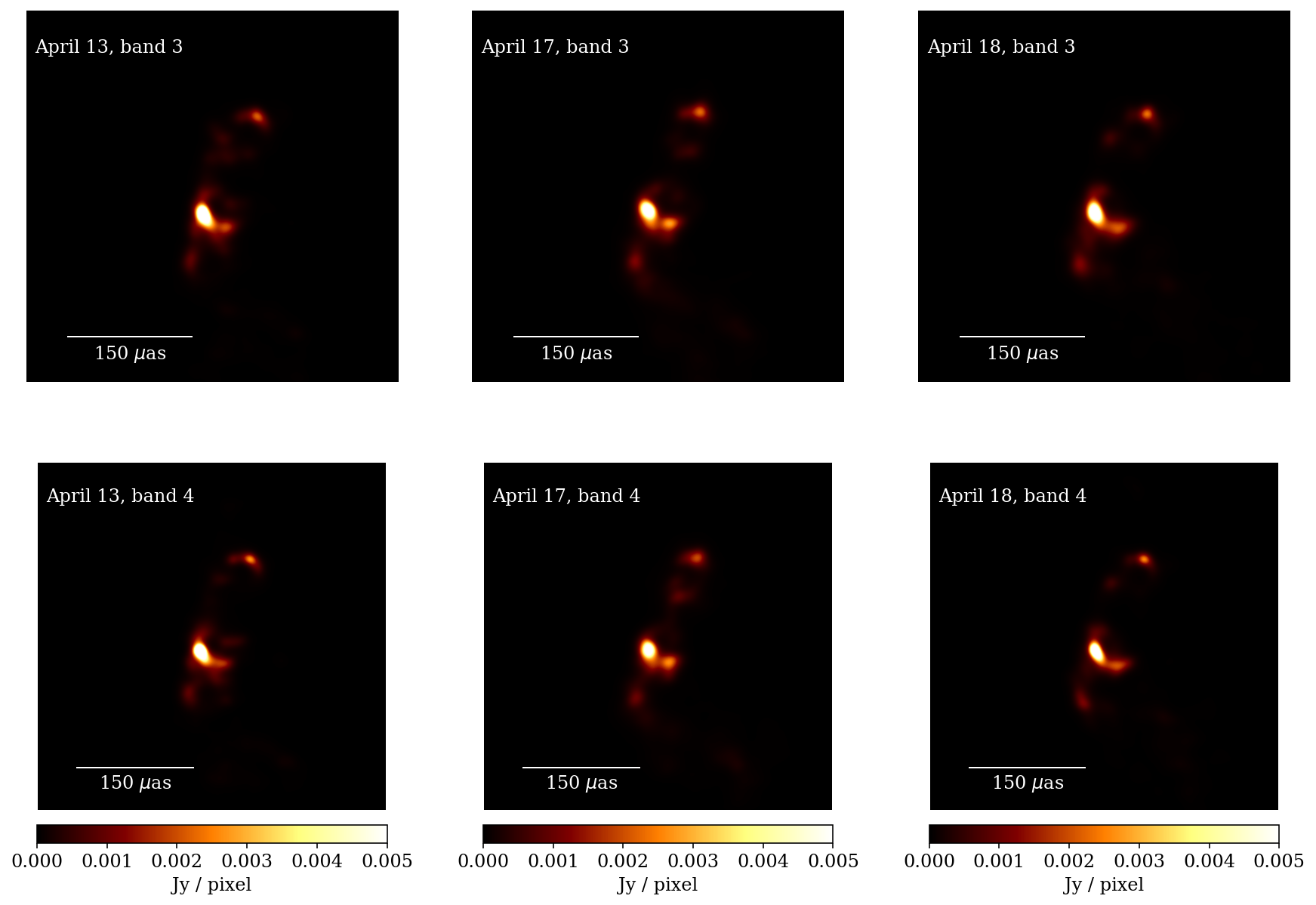}
    \caption{Recovered images (Comrade).}
    \label{fig:images}
\end{figure}

\begin{figure}
    \centering
    \includegraphics[width=\linewidth]{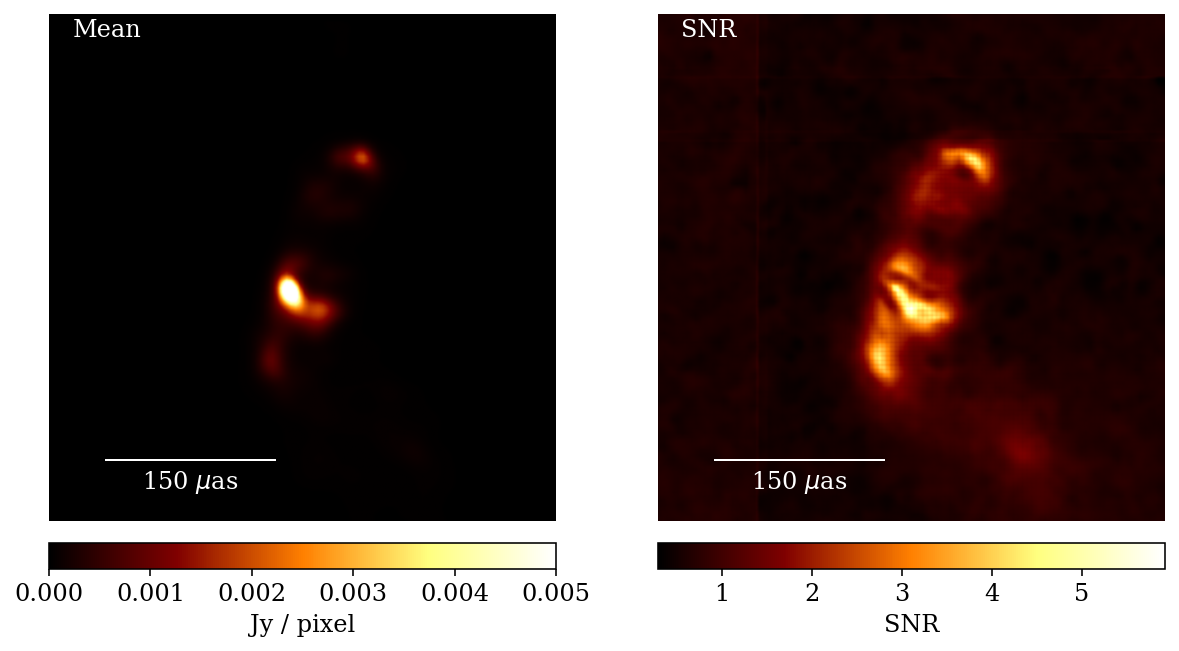}
    \caption{Left panel: Average of the images recovered by \texttt{Comrade} across all three days and both bands. Right panel: S/N in image domain derived from 500 posterior samples from the \texttt{Comrade} posterior. The pixel size is $1\,\mu\mathrm{as}^2$.}
    \label{fig:comrade_uncertainty}
\end{figure}

\begin{figure}
    \centering
    \includegraphics[width=\linewidth]{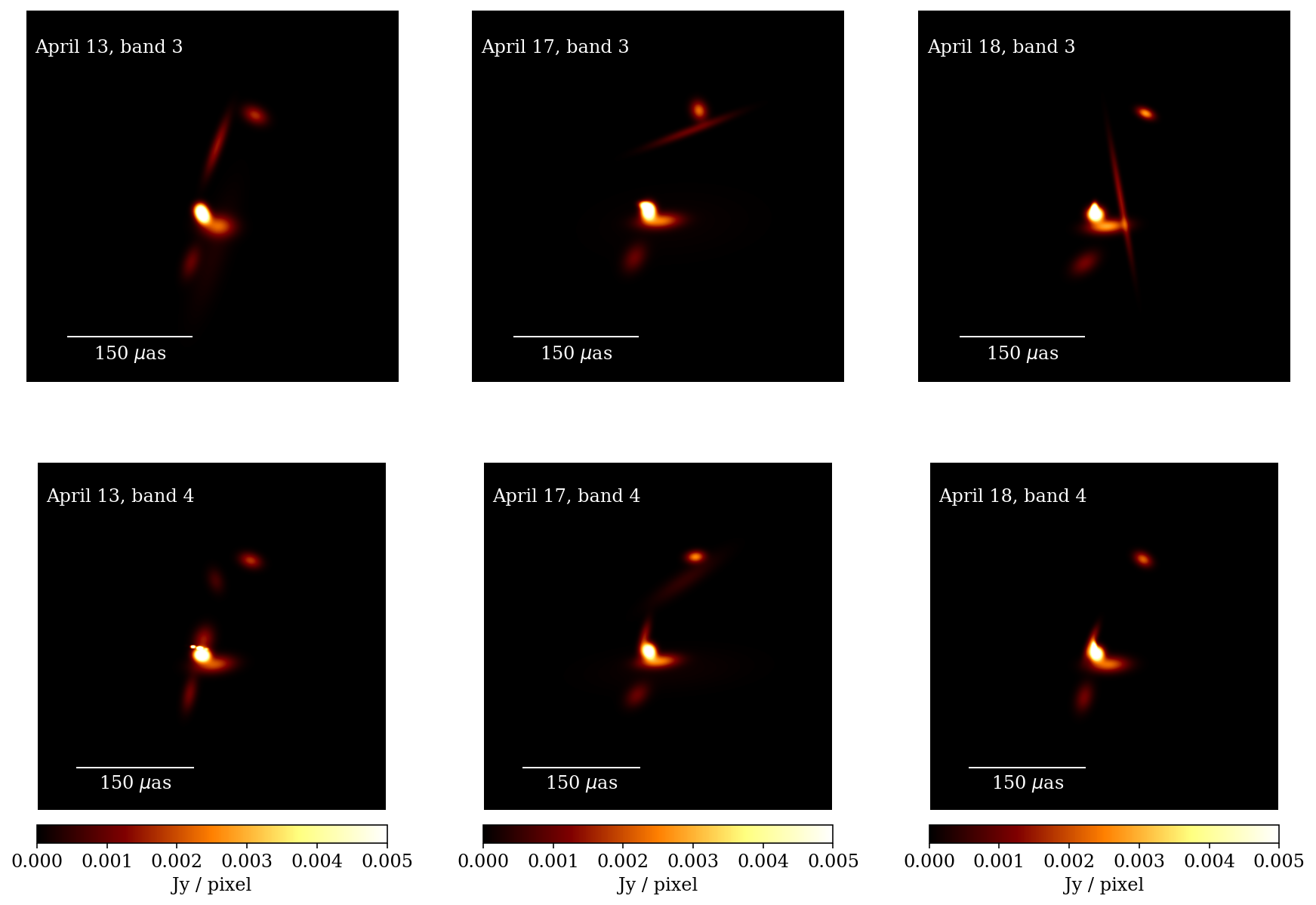}
    \caption{Model fitting results.}
    \label{fig:models}
\end{figure}

\begin{figure}
    \centering
    \includegraphics[width=\linewidth]{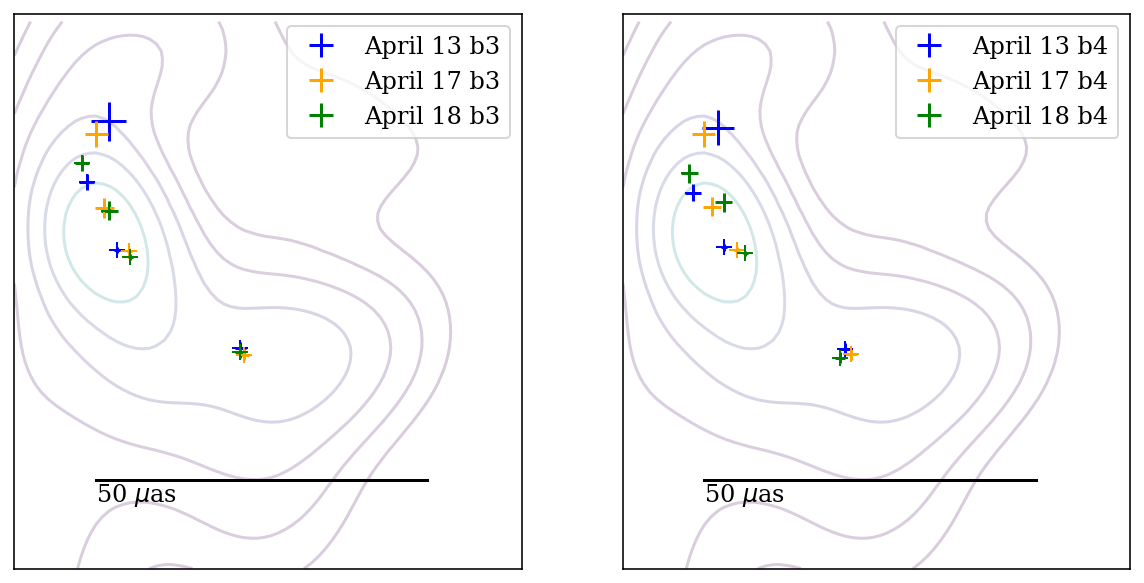}
    \caption{Location of the C2-x model-fitted components for band 3 (left) and band 4 (right). The blue, orange, and green crosses mark the positions at the three different observing epochs. The marker size indicates the model statistical fit uncertainty derived from half the beam size divided by the median S/N. The contours correspond to the 230\,GHz image reconstructed with \texttt{Comrade}. The two panels display different zoom levels of the image. The contour levels increase by a factor of 2 from $1.56\%$ to $50\%$ of the respective peak total intensity. All maps and components shown were aligned on C0.}
    \label{fig:modelling_bands}
\end{figure}

\begin{figure}
    \centering
    \includegraphics[width=\linewidth]{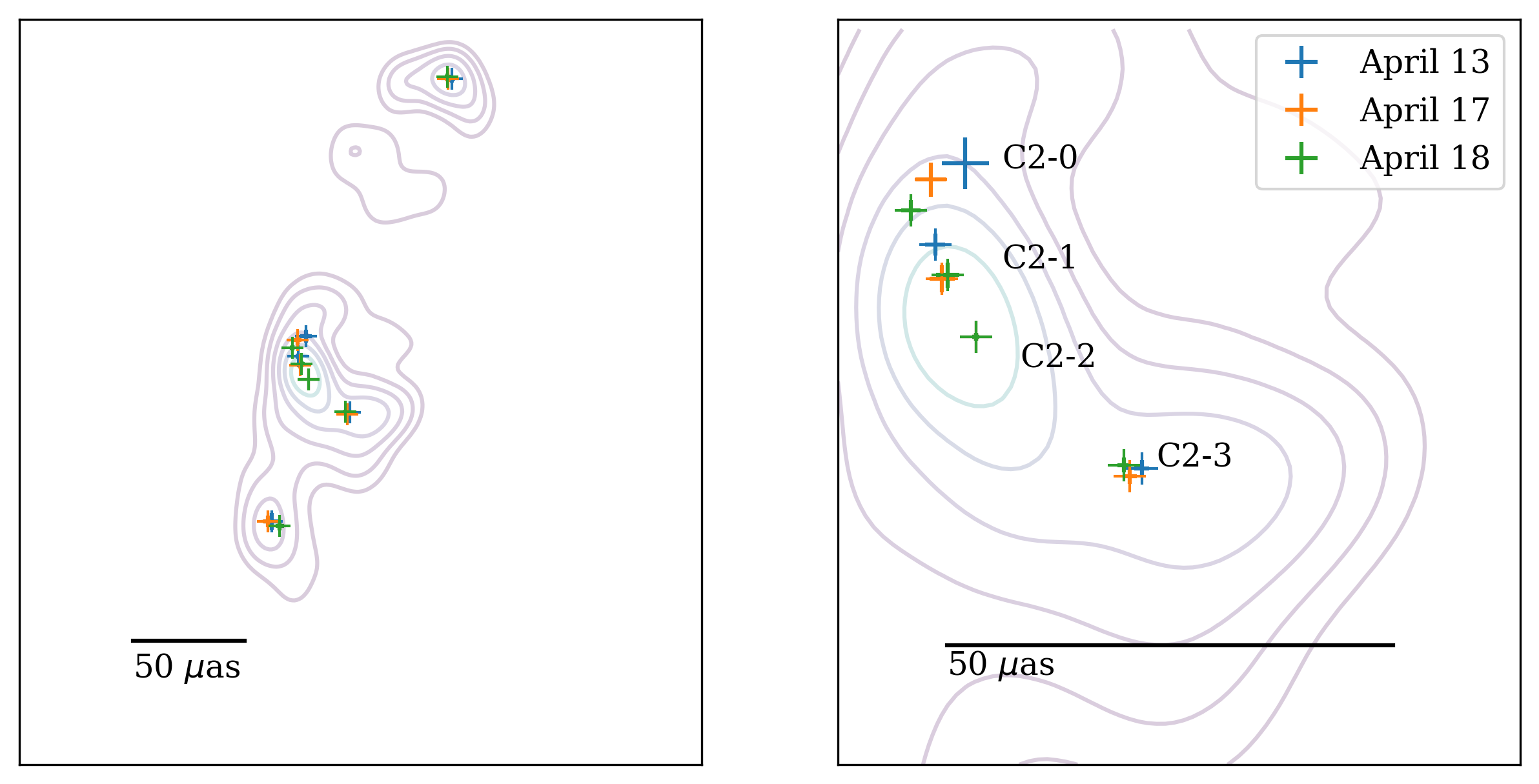}
    \caption{Model-fitting components when aligned on C2-2. The contour levels increase by a factor of 2 from $1.56\%$ to $50\%$ of the respective peak flux.}
    \label{fig:components_alternative_alignment}
\end{figure}

\begin{figure*}[ht]
    \centering
    \begin{subfigure}[b]{0.32\textwidth}
        \centering
        \includegraphics[width=\linewidth]{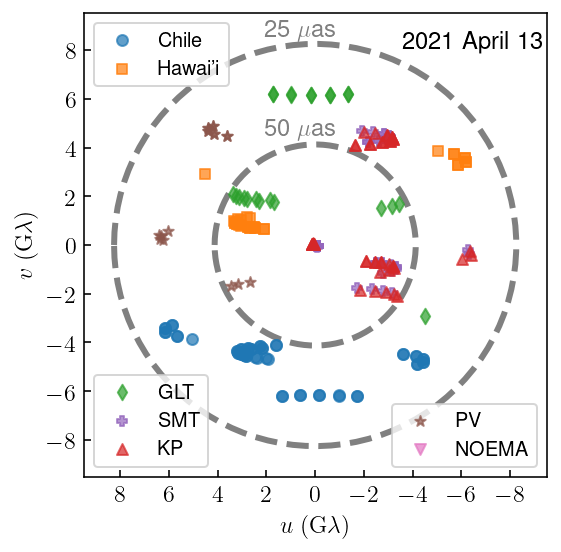}
        \caption{April 13.}
    \end{subfigure}
    \hfill
    \begin{subfigure}[b]{0.32\textwidth}
        \centering
        \includegraphics[width=\linewidth]{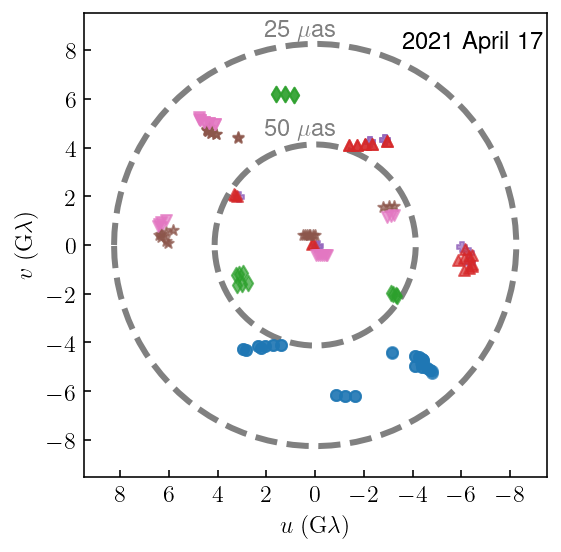}
        \caption{April 17.}
    \end{subfigure}
    \hfill
    \begin{subfigure}[b]{0.32\textwidth}
        \centering
        \includegraphics[width=\linewidth]{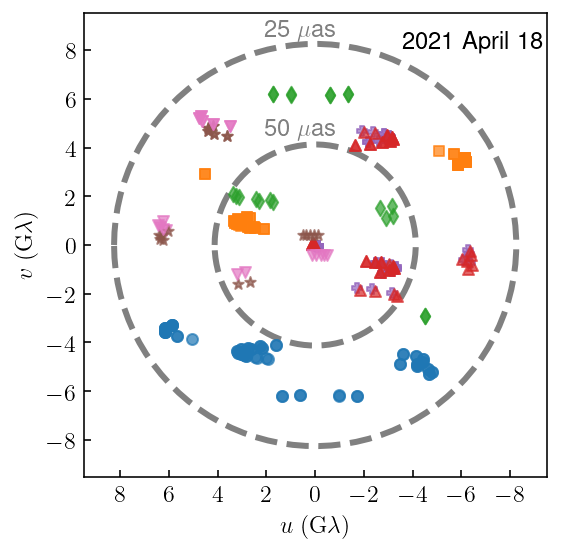}
        \caption{April 18.}
    \end{subfigure}
    \caption{U-V coverage of 3C~279 for the three 2021 EHT observing epochs.}
    \label{fig:uvcov}
\end{figure*}

Additional \texttt{Comrade} reconstruction for the 3 epochs and 2 bands are shown in \autoref{fig:images}. The average across these images, as well as with its statistical significance, is shown in Fig. \ref{fig:comrade_uncertainty}. Model fitting results obtained with \difmap are shown in \autoref{fig:models}. We compare the modeling results of the C2-x components for band 3 and band 4 in Fig. \ref{fig:modelling_bands}. We note the good match between the results obtained for band 3 and for band 4. Moreover, we show the model fit results when components are aligned with component C2-2 in \autoref{fig:components_alternative_alignment}. The u-v-coverage for the three 2021 observing epochs is shown in \autoref{fig:uvcov}.

\begin{figure*}
    \section{\texttt{kine} reconstruction of 3C~279 at 43\,GHz} \label{app:kine iamges}
    We show \texttt{kine} reconstructions across different 4 observing epochs in April 2021 in \autoref{fig:43GHz}. The images show a evolving structure, which in all cases shows a northern extension consistent with the component C0 identified in the 2021 EHT data.
    \vspace{0.5em}

    \centering
    \includegraphics[width=\textwidth]{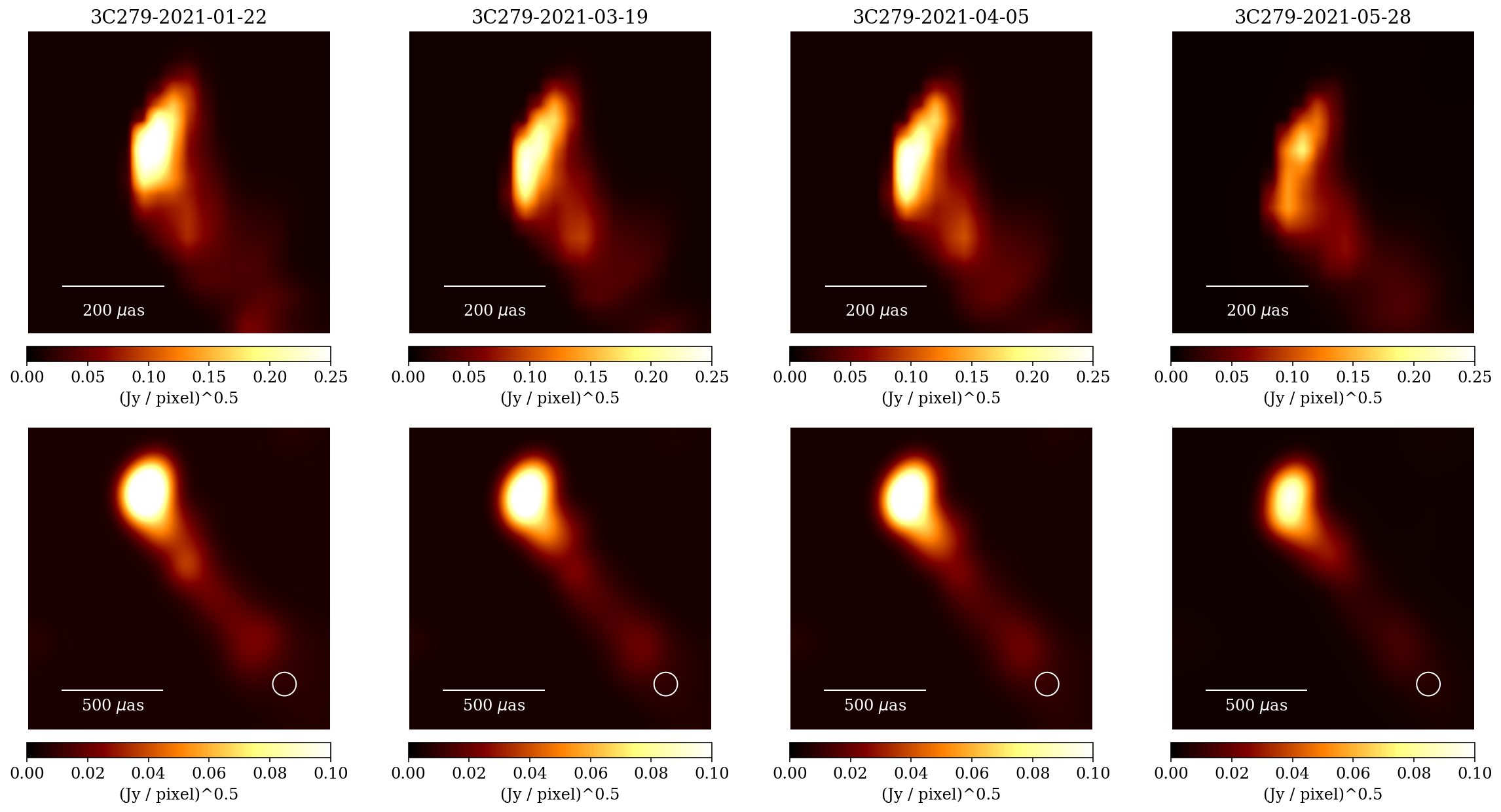}
    \caption{43\,GHz images of 3C~279 as obtained by the University of Boston BEAM-ME program. The reconstructions were performed with the dynamic imaging technique \texttt{kine}. Top: Epochs taken around April 2021 at native resolution. Bottom: Epochs convolved with a Gaussian with the instrumental resolution of $116\,\mu\mathrm{as}$.}
    \label{fig:43GHz}
\end{figure*}

\begin{table*}[tb]

\section{Table of image components}
\vspace{0.5em}

\centering
\small
\caption{Image components and fitted elliptical Gaussian parameters.}
\label{tab:components}

\setlength{\tabcolsep}{3pt}
\begin{tabular}{llccccccccc}
\hline\hline
Day & Band &
Component &
Flux [Jy] &
RA [$\mu$as] &
Dec [$\mu$as] &
Maj. [$\mu$as] &
Min. [$\mu$as] &
PA [$^\circ$] &
$T_B$ [$10^{10}$ K] \\
\hline

& & C0   & $0.659 \pm 0.120$ & $0.0 \pm 0.0$ & $0.0 \pm 0.0$ & $20.8 \pm 0.19$ & $14.9 \pm 0.19$ & $69.5 \pm 1.5$ & $4.91 \pm 0.08$ \\
& & C1   & $0.920 \pm 0.136$ & $42.3 \pm 0.19$ & $-29.2 \pm 0.19$ & $51.1 \pm 0.19$ & $12.0 \pm 0.19$ & $-2.2 \pm 0.56$ & $3.45 \pm 0.06$ \\
& & C2-0 & $0.604 \pm 0.117$ & $64.3 \pm 0.19$ & $-103.5 \pm 0.19$ & $31.3 \pm 0.19$ & $31.3 \pm 0.19$ & $0.0 \pm 0.0$ & $1.42 \pm 0.01$ \\
April 13 & 3 & C2-1 & $0.448 \pm 0.110$ & $67.6 \pm 0.19$ & $-111.7 \pm 0.19$ & $14.5 \pm 0.19$ & $6.6 \pm 0.19$ & $-77.6 \pm 0.78$ & $10.82 \pm 0.35$ \\
& & C2-2 & $1.670 \pm 0.195$ & $63.1 \pm 0.19$ & $-120.9 \pm 0.19$ & $13.8 \pm 0.19$ & $12.4 \pm 0.19$ & $9.4 \pm 4.35$ & $22.53 \pm 0.47$ \\
& & C2-3 & $1.599 \pm 0.189$ & $44.8 \pm 0.19$ & $-134.2 \pm 0.19$ & $32.8 \pm 0.19$ & $20.8 \pm 0.19$ & $86.5 \pm 1.17$ & $5.41 \pm 0.06$ \\
& & C3   & $0.424 \pm 0.109$ & $79.3 \pm 0.19$ & $-178.1 \pm 0.19$ & $28.2 \pm 0.19$ & $28.2 \pm 0.19$ & $0.0 \pm 0.0$ & $1.23 \pm 0.01$ \\
\hline

& & C0   & $0.685 \pm 0.121$ & $0.0 \pm 0.0$ & $0.0 \pm 0.0$ & $18.1 \pm 0.19$ & $15.1 \pm 0.19$ & $76.5 \pm 2.59$ & $5.77 \pm 0.10$ \\
& & C1   & $0.859 \pm 0.132$ & $18.5 \pm 0.19$ & $-27.2 \pm 0.19$ & $65.4 \pm 0.19$ & $29.6 \pm 0.19$ & $-53.9 \pm 0.78$ & $1.02 \pm 0.01$ \\
& & C2-0 & $0.592 \pm 0.116$ & $66.2 \pm 0.19$ & $-105.2 \pm 0.19$ & $20.2 \pm 0.19$ & $20.2 \pm 0.19$ & $0.0 \pm 0.0$ & $3.35 \pm 0.05$ \\
April 13 & 4 & C2-1 & $0.424 \pm 0.109$ & $64.9 \pm 0.19$ & $-115.2 \pm 0.19$ & $11.6 \pm 0.19$ & $11.6 \pm 0.19$ & $0.0 \pm 0.0$ & $7.33 \pm 0.17$ \\
& & C2-2 & $1.612 \pm 0.190$ & $61.2 \pm 0.19$ & $-121.0 \pm 0.19$ & $15.2 \pm 0.19$ & $10.8 \pm 0.19$ & $-0.7 \pm 1.48$ & $22.63 \pm 0.50$ \\
& & C2-3 & $1.708 \pm 0.198$ & $44.3 \pm 0.19$ & $-135.1 \pm 0.19$ & $38.9 \pm 0.19$ & $17.9 \pm 0.19$ & $-89.2 \pm 0.79$ & $5.67 \pm 0.07$ \\
& & C3   & $0.601 \pm 0.117$ & $78.9 \pm 0.19$ & $-178.2 \pm 0.19$ & $27.3 \pm 0.19$ & $27.3 \pm 0.19$ & $0.0 \pm 0.0$ & $1.86 \pm 0.02$ \\
\hline

& & C0   & $0.567 \pm 0.115$ & $0.0 \pm 0.0$ & $0.0 \pm 0.0$ & $16.9 \pm 0.19$ & $12.3 \pm 0.19$ & $30.2 \pm 1.58$ & $6.29 \pm 0.12$ \\
& & C1   & $0.962 \pm 0.139$ & $-18.3 \pm 0.19$ & $-36.0 \pm 0.19$ & $117.5 \pm 0.19$ & $117.5 \pm 0.19$ & $0.0 \pm 0.0$ & $0.161 \pm 0.001$ \\
& & C2-0 & $0.687 \pm 0.121$ & $68.3 \pm 0.19$ & $-109.1 \pm 0.19$ & $30.9 \pm 0.19$ & $30.9 \pm 0.19$ & $0.0 \pm 0.0$ & $1.66 \pm 0.22$ \\
April 17 & 3 & C2-1 & $0.458 \pm 0.110$ & $64.2 \pm 0.19$ & $-115.6 \pm 0.19$ & $15.0 \pm 0.19$ & $9.2 \pm 0.19$ & $-5.1 \pm 1.11$ & $7.70 \pm 0.19$ \\
& & C2-2 & $1.618 \pm 0.190$ & $61.1 \pm 0.19$ & $-121.8 \pm 0.19$ & $14.0 \pm 0.19$ & $12.1 \pm 0.19$ & $-35.3 \pm 3.19$ & $22.11 \pm 0.47$ \\
& & C2-3 & $1.717 \pm 0.199$ & $44.8 \pm 0.19$ & $-134.7 \pm 0.19$ & $35.1 \pm 0.19$ & $18.0 \pm 0.19$ & $-89.3 \pm 0.88$ & $6.24 \pm 0.08$ \\
& & C3   & $0.598 \pm 0.116$ & $73.8 \pm 0.19$ & $-180.8 \pm 0.19$ & $21.5 \pm 0.19$ & $21.5 \pm 0.19$ & $0.0 \pm 0.0$ & $2.99 \pm 0.04$ \\
\hline

& & C0   & $0.668 \pm 0.120$ & $0.0 \pm 0.0$ & $0.0 \pm 0.0$ & $23.1 \pm 0.19$ & $11.6 \pm 0.19$ & $81.5 \pm 0.86$ & $5.76 \pm 0.11$ \\
& & C1   & $0.820 \pm 0.129$ & $28.9 \pm 0.19$ & $-30.7 \pm 0.19$ & $81.4 \pm 0.19$ & $38.7 \pm 0.19$ & $-82.0 \pm 0.81$ & $0.60 \pm 0.00$ \\
& & C2-0 & $0.693 \pm 0.122$ & $64.3 \pm 0.19$ & $-104.4 \pm 0.19$ & $33.0 \pm 0.19$ & $33.0 \pm 0.19$ & $0.0 \pm 0.0$ & $1.47 \pm 0.01$ \\
April 17 & 4 & C2-1 & $0.475 \pm 0.111$ & $67.9 \pm 0.19$ & $-113.2 \pm 0.19$ & $12.5 \pm 0.19$ & $9.9 \pm 0.19$ & $-8.2 \pm 2.09$ & $8.87 \pm 0.22$ \\
& & C2-2 & $1.634 \pm 0.192$ & $63.3 \pm 0.19$ & $-120.6 \pm 0.19$ & $14.6 \pm 0.19$ & $12.5 \pm 0.19$ & $17.4 \pm 2.98$ & $20.55 \pm 0.42$ \\
& & C2-3 & $1.709 \pm 0.198$ & $45.4 \pm 0.19$ & $-134.3 \pm 0.19$ & $39.5 \pm 0.19$ & $19.6 \pm 0.19$ & $87.7 \pm 0.85$ & $5.09 \pm 0.06$ \\
& & C3   & $0.499 \pm 0.112$ & $76.9 \pm 0.19$ & $-175.0 \pm 0.19$ & $28.5 \pm 0.19$ & $28.5 \pm 0.19$ & $0.0 \pm 0.0$ & $1.42 \pm 0.01$ \\
\hline

& & C0   & $0.681 \pm 0.121$ & $0.0 \pm 0.0$ & $0.0 \pm 0.0$ & $18.6 \pm 0.19$ & $14.7 \pm 0.19$ & $78.9 \pm 2.01$ & $5.74 \pm 0.10$ \\
& & C1   & $0.858 \pm 0.132$ & $16.8 \pm 0.19$ & $-27.0 \pm 0.19$ & $66.5 \pm 0.19$ & $29.2 \pm 0.19$ & $-55.9 \pm 0.76$ & $1.02 \pm 0.01$ \\
& & C2-0 & $0.591 \pm 0.116$ & $66.3 \pm 0.19$ & $-105.2 \pm 0.19$ & $20.6 \pm 0.19$ & $20.6 \pm 0.19$ & $0.0 \pm 0.0$ & $3.20 \pm 0.04$ \\
April 18 & 3 & C2-1 & $0.425 \pm 0.109$ & $65.1 \pm 0.19$ & $-115.0 \pm 0.19$ & $11.4 \pm 0.19$ & $11.4 \pm 0.19$ & $0.0 \pm 0.0$ & $7.53 \pm 0.18$ \\
& & C2-2 & $1.611 \pm 0.190$ & $61.4 \pm 0.19$ & $-120.9 \pm 0.19$ & $15.0 \pm 0.19$ & $11.3 \pm 0.19$ & $-0.3 \pm 1.73$ & $22.07 \pm 0.48$ \\
& & C2-3 & $1.710 \pm 0.198$ & $44.5 \pm 0.19$ & $-134.9 \pm 0.19$ & $38.6 \pm 0.19$ & $18.5 \pm 0.19$ & $-89.2 \pm 0.82$ & $5.52 \pm 0.06$ \\
& & C3   & $0.602 \pm 0.117$ & $78.6 \pm 0.19$ & $-178.2 \pm 0.19$ & $27.6 \pm 0.19$ & $27.6 \pm 0.19$ & $0.0 \pm 0.0$ & $1.82 \pm 0.02$ \\
\hline

& & C0   & $0.529 \pm 0.113$ & $0.0 \pm 0.0$ & $0.0 \pm 0.0$ & $18.0 \pm 0.19$ & $7.9 \pm 0.19$ & $66.6 \pm 0.77$ & $8.55 \pm 0.23$ \\
& & C1   & $0.862 \pm 0.132$ & $-14.6 \pm 0.19$ & $-29.0 \pm 0.19$ & $126.2 \pm 0.19$ & $126.2 \pm 0.19$ & $0.0 \pm 0.0$ & $0.125 \pm 0.0003$ \\
& & C2-0 & $0.758 \pm 0.125$ & $68.5 \pm 0.19$ & $-110.5 \pm 0.19$ & $18.9 \pm 0.19$ & $18.9 \pm 0.19$ & $0.0 \pm 0.0$ & $4.91 \pm 0.07$ \\
April 18 & 4 & C2-1 & $0.507 \pm 0.112$ & $63.3 \pm 0.19$ & $-114.4 \pm 0.19$ & $14.4 \pm 0.19$ & $11.6 \pm 0.19$ & $9.2 \pm 2.18$ & $7.04 \pm 0.15$ \\
& & C2-2 & $1.672 \pm 0.195$ & $60.2 \pm 0.19$ & $-121.4 \pm 0.19$ & $13.7 \pm 0.19$ & $12.2 \pm 0.19$ & $-31.2 \pm 3.78$ & $23.01 \pm 0.49$ \\
& & C2-3 & $1.686 \pm 0.196$ & $46.2 \pm 0.19$ & $-135.4 \pm 0.19$ & $33.9 \pm 0.19$ & $18.2 \pm 0.19$ & $88.3 \pm 0.92$ & $6.30 \pm 0.08$ \\
& & C3   & $0.594 \pm 0.116$ & $73.1 \pm 0.19$ & $-181.0 \pm 0.19$ & $22.0 \pm 0.19$ & $22.0 \pm 0.19$ & $0.0 \pm 0.0$ & $2.82 \pm 0.04$ \\
\hline

\end{tabular}
\end{table*}

\end{document}